\documentclass[usenatbib]{mn2e}
	
\usepackage{graphicx}
\usepackage[space]{grffile}
\usepackage{latexsym}
\usepackage{textcomp}
\usepackage{longtable}
\usepackage{multirow,booktabs}
\usepackage{amsfonts,amsmath,amssymb}
\usepackage{natbib}
\usepackage{url}
\usepackage{hyperref}
\usepackage{deluxetable}
\usepackage[english]{babel}

\newcommand{\apj}{ApJ}

\newcommand{\mnras}{MNRAS}

\newcommand{\nat}{Nature}

\usepackage{xspace}

\def \Tvir {$T_{\mathrm{vir}}$\xspace}

\def \kms {km\ s$^{-1}$\xspace}
\def \Msol {$\mathrm{M_\odot}$\xspace}  
\def \Mpeak {$M_{\mathrm{peak}}$\xspace}
\def \Mstar {$M_{\mathrm{\star}}$\xspace}

\def \hMpc {$h^{-1}$\ Mpc\xspace}

\def \Mvir {$M_{\mathrm{vir}}$\xspace}

\def \zre {$z_{\mathrm{re}}$\xspace}
\def \Tvir {$T_{\mathrm{vir}}$\xspace}
\def \Vmax {$V_{\mathrm{max}}$\xspace}

\def \Mpeak {$M_{\mathrm{peak}}$\xspace}

\def \rockstar {{\sc{rockstar}}\xspace}

\def \caterpillar {\textit{Caterpillar}\xspace}
\def \caterpillarproject {\textit{Caterpillar\ Project}\xspace}

\title[Tracing the first stars and galaxies of the Milky Way]{Tracing the first stars and galaxies of the Milky Way}

\author[B. F. Griffen et al.]{\parbox{17cm}{Brendan F. Griffen$^{1}$\thanks{e-mail: brendan.f.griffen@gmail.com},
Gregory A. Dooley$^{1}$,
Alexander P. Ji$^{1}$,
Brian W. O'Shea$^{2,3}$,
Facundo A. G\'{o}mez$^{4}$,
Anna Frebel$^{1}$
}\vspace{0.3cm}\\
$^{1}$Department of Physics, Kavli Institute for Astrophysics and Space Research, Massachusetts Institute of Technology, \\ Cambridge, MA 02139, USA\\
$^{2}$Department of Physics and Astronomy,  Department of Computational Mathematics, Science and Engineering, \\ \ \ and National Superconducting Cyclotron Laboratory, Michigan State University, East Lansing, MI 48824, USA\\
$^{3}$Joint Institute for Nuclear Astrophysics, Center for Evolution of the Elements, East Lansing, MI 48824, USA\\
$^{4}$Max-Planck-Institut f\"ur Astrophysik, 85748 Garching, Germany}

\begin{document}

\date{Submitted to MNRAS: November 2, 2016.}

\maketitle

\label{firstpage}

\begin{abstract}
We use 30 high-resolution dark matter halos of the $\caterpillar$ simulation suite to probe the first stars and galaxies of Milky Way-mass systems. We quantify the environment of the high-$z$ progenitors of the Milky Way and connect them to the properties of the host and satellites today. We identify the formation sites of the first generation of Population\,III (Pop\,III) stars ($z$ $\sim$ 25) and first galaxies ($z$ $\sim$ 22) with several different models based on a minimum halo mass including a simple model for Lyman-Werner feedback. Through this method we find approximately 23,000 $\pm$ 5,000 Pop\,III potentially star-forming sites per Milky Way-mass host, though this number is drastically reduced to $\sim$550 star-forming sites when Lyman-Werner feedback is included, as it has critical effects at these length scales. The majority of these halos identified form in isolation (96$\%$ at $z$ = 15) and are not subject to external enrichment by neighboring halos (median separation $\sim$1\,pkpc at $z$ = 15), though half merge with a system larger than themselves within 1.5 Gyrs. Approximately 55$\%$ of the entire population has merged into the host halo by $z$ = 0. Using particle tagging, we additionally trace the Pop\, III remnant population to $z$ = 0 and find an order of magnitude scatter in their number density at small (i.e.\ r $<$ 5\ kpc) and large (i.e.\ r $>$ 50\ kpc) galactocentric radii at $z$ = 0. Using our large number of realizations, we provide fitting functions for determining the number of progenitor minihalo and atomic cooling halo systems that present-day dwarf galaxies and the Magellanic cloud system might have accreted since their formation. We demonstrate that observed dwarf galaxies with stellar masses below 10$^{4.6}$\ \Msol are unlikely to have merged with any other star-forming systems.

\end{abstract}


\section{Introduction}
The epoch of the first stars and first galaxies remains a poorly understood period in the Universe's history, although it is broadly known how the first billion years unfolded. Following recombination (z$\sim$1100), small scale density fluctuations collapsed into dark matter halos containing gaseous material capable of molecular hydrogen (H$_2$) cooling. Once gas densities in these ``minihalos'' were sufficiently high, the first stars, Population\,III (Pop\,III), were able to form, thus marking the end of the so-called ``Dark Ages'' \citep{Tegmark:1997ce}.

These Pop\,III stars were predominantly massive (e.g.\ \citealt{1999ApJ...527L...5B}) and thus exploded soon after formation as supernovae (SNe). Their deaths produced vast quantities of ionizing radiation and metals, impacting the conditions for subsequent star formation. 
The metals may have ``cross-polluted'' nearby minihalos (e.g. \citealt{Whalen:2008hu,Smith:2015dw}), and in the case of pair-instability supernovae (PISNe) enriched vast volumes of the early inter-galactic medium (IGM) \citep{2003Natur.422..871U,Whalen:2004eq, Kitayama:2005ea}.

This enriched and ionized environment set a blueprint for more massive galaxies which assembled soon after \citep{Greif:2007ha,Wise:2008bq,OShea:2015fz}. As structure formation progressed ($z$ $\sim$ 25, \citealt{Greif:2008ke}), these more massive dark matter halos (10$^8$\,\Msol) formed with gas that could cool via atomic hydrogen (\Tvir$\sim$$10^4$\,K) and are widely regarded as the ``first galaxies,'' heralding the first period of galaxy formation \citep{Bromm:2011dh}. While remarkable progress has been made in arriving at this broad picture of early structure formation, many of the details of both first star and first galaxy formation and associated chemical and physical processes remain elusive (see \citealt{Frebel:2015kk} for a review).

Observational access to these critical periods is quite limited. The optical depth to reionization derived from the cosmic microwave background provides a global constraint on reionization . Other studies have used very deep images of high-z sources (e.g.\ \citet{2015ApJ...810...71F, 2015ApJ...808..139S}) or absorption from high-z quasars \citep[e.g.][]{2015PASA...32...45B} to study the brightest objects at $6 \lesssim z \lesssim 10$. Future data from the \textit{James Webb Space Telescope} \citep{2006SSRv..123..485G} or 21 cm cosmology \citep{2012RPPh...75h6901P} will provide valuable further constraints. 

One of the best ways to study this early period is by examining local objects. For instance, star formation histories of local group galaxies, and chemical abundances of metal-poor stars in the Milky Way stellar halo (stellar archaeology) or in its satellite dwarf galaxies (dwarf galaxy archaeology). A key step in interpreting these local observations is developing a principled method of connecting low-z stellar systems to their high-z progenitors.

Cosmological simulations have been used extensively to study the non-linear regime of structure formation, but few have been able to resolve and follow the smallest building blocks, which formed in the early universe, to the present day. Indeed, it is still not well understood how many progenitor systems made up the Milky Way nor do we know where they formed and, perhaps most importantly, where any can be found today if they survived (see \citealt{Frebel:2010kw} for a review). This is primarily due to the high redshift universe  being studied from only two vantage points, either (a) moderately large nondescript volumes (e.g.\ \citealt{Ricotti00, Ishiyama:2013hb,Ishiyama:2016vt,Sasaki:2014fva,OShea:2015fz}) or (b) small volumes encapsulating single host halos at extremely high-resolution (e.g. \citealt{Smith:2015dw,Stacy:2016ux}). By virtue of the computational cost of carrying out hydrodynamical simulations at the required resolutions to resolve the first stellar systems, these calculations usually only run to relatively high redshifts ($z$ $\sim$ 10), prohibiting our ability to test them against local observational data. Currently, exclusively dark matter-only simulations are capable to resolve both the minihalo progenitors ($z$ $>$ 15) of the Milky Way \textit{and} to trace their evolution to $z$ = 0 (e.g.\ \citealt{Springel:2008gd,Diemand:2008hr,Griffen:2016kn}). Until the required hardware and hydrodynamic codes with the necessary physics become available, minihalos and atomic cooling halos have to be modeled semi-analytically (i.e.\ using the halo properties derived from halo finders to determine the nature of their gas and stellar content).

Indeed, there have been a number of works which have made attempts to connect the high-z universe to the present day via either semi-analytical methods or direct N-body simulations. All of them, however, suffer from at least one drawback, either (a) they contain no spatial information about where the high-z star forming halos reside today (e.g.\ \citealt{2015MNRAS.447.3892H}, \citealt{Smith:2015dw}), or (b) they do not model the critical influence of Lyman-Werner (LW) feedback on the first stellar systems (e.g.\ \citealt{Gao:2010cva}), or (c) they lack statistical power to investigate halo-to-halo scatter (e.g.\ \citealt{Tumlinson:2009ia,Bovill:2011iz,Corlies:2013bs,Ishiyama:2016vt}).

In this paper, we address these three issues directly by exploiting the high-resolution cosmological dark matter simulations of the \caterpillarproject (\href{http://www.caterpillarproject.org}{www.caterpillarproject.org}, \citealt{Griffen:2016kn}). Specifically, this work has the following properties which combined is the first of its kind:  (a) spatial information about the formation sites and their subsequent evolution to $z$ = 0 (b) a model for LW feedback on the first stellar systems and (c) 30 realizations allowing robust halo-to-halo variations to be studied. We adopt simple models to identify the sites of first star and first galaxy formation and include a toy model for chemical enrichment which allows us to separate halos with metal enrichment driven exogenously (externally) and endogenously (internally). We trace our candidate minihalos and first galaxy halos using their most bound particles to determine where their progenitors are today. This approach connects the high-$z$ star formation processes to surviving stars in low-$z$ environments today (e.g.\ dwarf galaxies and the halo), probes the building blocks of the Milky Way's metal-poor stellar halo, assists in the hunt for the surviving relics from a unique period of our Galaxy's assembly history, and informs how differing formation histories of similarly-sized galaxies can affect observable properties of metal-poor stellar populations. This is the first time that first star and first galaxy formation is  studied with such a wide sample of simulations within the context of the entire Milky Way host assembly.

In Section \ref{sec:method}, we describe our numerical simulation suite and method for identifying and tracking dark matter halos. In Section \ref{sec:sf}, we present our model for Pop\,III and Pop\, II star formation, including our method of treating the LW background. Our results are presented in Section \ref{sec:results} where we detail the clustered nature of the high-$z$ progenitor Milky Way and highlight how this critically impacts the present day abundance of possible surviving stellar populations. We additionally discuss these results in the context of the recent discovery of $r$-process enhanced metal-poor stars inside one of today's ultra-faint dwarf galaxies. Section \ref{sec:conclusions} provides our final concluding remarks and avenues for future work.

\section{Simulations}
\label{sec:method}

We analyze 30 of the dark matter only cosmological halos of the \caterpillarproject first presented in \citet{Griffen:2016kn}. Each of the 30 halos in our sample are similar in mass to that of the Milky Way and come from a somewhat isolated environment (no nearby clusters). The halos were identified from a larger parent simulation which followed the growth of structure in a periodic box of comoving length 100 \hMpc with 1024$^3$ particles ($m_p$ = 1.22 $\times$ 10$^7$ \Msol). For the underlying cosmological model we adopt the $\Lambda$CDM parameter set characterized by a \textit{Planck} 2013 cosmology given by $\Omega_m=0.32$, $\Omega_\Lambda=0.68$, $\Omega_b=0.05$, $n_s=0.96$, $\sigma_8=0.83$ and Hubble constant, H = 100 $h$ km s$^{-1}$ Mpc$^{-1}$ = 67.11 km s$^{-1}$ Mpc$^{-1}$ \citep{Collaboration:2014dt}. All initial conditions were constructed using {\sc{music}} (\citealt{Hahn:2011gj}).  We identify dark matter halos via a modified version of  \rockstar \citep{Behroozi:2013cn} and construct merger trees using {\sc{Consistent-Trees}} \citep{Behroozi:2012dz}. \rockstar assigns virial masses to halos, \Mvir, using the evolution of the virial relation from \cite{Bryan:1998cc} for our particular cosmology. At \textit{z} = 0, this definition corresponds to an over-density of 104 $\times$ the critical density of the Universe. The temporal resolution is $\sim$5 Myrs/snapshot to $z$ = 6 and $\sim$50 Myrs to $z$ = 0.

\caterpillar halos are zoom-in re-simulations of a parent volume. Particular care was taken to ensure that we restrict our study to only the high-resolution volume of the Milky Way at $z>10$ and that no halos were contaminated. Indeed in all simulations of this kind, there will be halos in the catalogues which contain lower resolution particle types, particularly near the fringe of the high-resolution region. These halos have poorly determined virial masses and internal velocity dispersions, so they are excluded from our analysis ($<$1$\%$ of the total halo population on average). None of these contaminated halos end up anywhere near the host of the central Milky Way-mass system at $z=0$.

The dark matter particle mass of the fiducial \caterpillar simulation suite is 2.99 $\times$ 10$^4$ \Msol, resolving halos with masses of 10$^6$ \Msol ($\sim$30 particles). Whilst properties such as the velocity dispersion are not converged at the resolution limit, the total mass of the system is reliably determined \citealt{Power:2013ws}. 

We carried out a convergence check of this assumption (see Appendix A) using an even higher resolution run with a particle mass of 3.73 $\times$ 10$^3$ \Msol.  We find that the total number of systems identified between our fiducial run and our ultra-high-resolution counterpart is convergent.

\begin{table}
\caption{{Properties of the 30 \caterpillar halos used in analysis}}
\begin{center}
\begin{tabular}{lccrcc}
\hline
Name & $M_{\mathrm{vir}}$$^a$ & $R_{\mathrm{vir}}$$^b$ & c$^c$ & $V_{\mathrm{max}}$$^d$ & $z_{0.5}$\ $^e$ \\ 
 & (\Msol) & (kpc) &  & (km/s) &  \\
\hline
Cat-1 & 1.559 & 306.381 & 7.491 & 169.760 & 0.894 \\
Cat-2 & 1.791 & 320.906 & 8.370 & 178.850 & 0.742 \\
Cat-3 & 1.354 & 292.307 & 10.168 & 172.440 & 0.802 \\
Cat-4 & 1.424 & 297.288 & 8.574 & 164.340 & 0.936 \\
Cat-5 & 1.309 & 289.081 & 12.108 & 176.400 & 0.564 \\
Cat-6 & 1.363 & 292.947 & 10.195 & 171.650 & 1.161 \\
Cat-8 & 1.702 & 315.466 & 13.505 & 198.560 & 1.516 \\
Cat-9 & 1.322 & 289.987 & 12.401 & 177.410 & 1.255 \\
Cat-10 & 1.323 & 290.116 & 11.715 & 174.990 & 1.644 \\
Cat-12 & 1.763 & 319.212 & 11.401 & 191.260 & 1.336 \\
Cat-13 & 1.164 & 277.938 & 12.850 & 171.220 & 1.161 \\
Cat-14 & 0.750 & 240.119 & 9.135 & 137.440 & 1.144 \\
Cat-15 & 1.505 & 302.788 & 8.984 & 174.120 & 1.144 \\
Cat-16 & 0.982 & 262.608 & 11.737 & 155.360 & 1.315 \\
Cat-17 & 1.319 & 289.800 & 12.765 & 179.060 & 1.846 \\
Cat-18 & 1.407 & 296.100 & 7.886 & 163.920  & 0.493 \\
Cat-19 & 1.174 & 278.771 & 10.467 & 164.730 & 1.541 \\
Cat-20 & 0.762 & 241.387 & 13.376 & 149.150 & 1.492 \\
Cat-21 & 1.882 & 326.206 & 10.618 & 190.680 & 1.126 \\
Cat-22 & 1.495 & 302.114 & 10.666 & 180.650 & 0.841 \\
Cat-23 & 1.608 & 309.525 & 12.489 & 190.710 & 1.161 \\
Cat-24 & 1.334 & 290.867 & 11.378 & 176.910 & 1.144 \\
Cat-25 & 1.648 & 312.153 & 12.970 & 191.690 & 1.126 \\
Cat-26 & 1.018 & 265.828 & 8.130 & 147.960 & 0.555 \\
Cat-27 & 1.357 & 292.557 & 7.035 & 159.730 & 0.719 \\
Cat-29 & 1.594 & 308.698 & 10.646 & 182.810 & 0.980 \\
Cat-31 & 1.678 & 313.967 & 12.461 & 191.710 & 1.516 \\
Cat-33 & 1.675 & 313.855 & 13.322 & 197.710 & 1.878 \\
Cat-36 & 1.974 & 331.521 & 10.282 & 191.890 & 0.966 \\
Cat-37 & 1.848 & 324.250 & 12.854 & 197.950 & 1.492 \\
\hline
\end{tabular}
\end{center}
$a$: Halo virial mass based on \cite{Bryan:1998cc}. \\
$b$: Halo virial radius based on \cite{Bryan:1998cc}. \\
$c$: Concentration defined by ratio of the virial radius and the scale radius; $R_{\mathrm{vir}}/R_{\mathrm{s}}$. \\
$d$: Maximum of the halo's circular velocity. \\
$e$: Redshift at which half the mass of the host has formed. \\
\end{table}

\section{Modelling The Sites Of High-Redshift Star Formation}
\label{sec:sf}

To determine which dark matter halos host stellar material and later accrete into the Milky Way, we must consider the nature of star formation in the early Universe. Here, we take a simple approach to modelling star formation sites based on more detailed theoretical work.

Structure formation within $\Lambda$CDM proceeded first within small dark matter halos forming at early times and merging into larger halos. There are two periods which are significant for star formation at these early times and they both relate to the cooling mechanisms in metal-poor gas. The first of these periods is when star formation proceeds within dark matter halos of mass $\sim$10$^6$ \Msol, in which molecular hydrogen cooling is dominant (e.g. \citealt{Tegmark:1997ce}). The second important period of star formation occurs when the gas within larger halos of mass $\sim$10$^8$ \Msol are able to cool via atomic line cooling \citep{Oh:2002ki}. In the following two sections we outline how we model these two periods of first generation (Pop\,III) and second generation (Pop\,II) star formation.

\subsection{H$_2$ Cooling}

Pop\,III stars are by definition metal-free, and as such can only form in a minihalo with sufficient H$_2$ at the appropriate temperature and density to become gravitationally unstable and collapse \citep{Tegmark:1997ce, OShea:2007iy}. We assume the gas is in virial equilibrium with the dark matter halo so we can infer the gas temperature from the dark matter virial mass. The minimum temperature required for H$_2$ cooling to cause gas collapse \citep{Tegmark:1997ce} thus corresponds to a minimum halo mass that determines possible sites of Pop\,III star formation.

We identify halos in our merger tree when they first grow above the minimum threshold for collapse. We additionally ensure that none of the progenitors on any branch that merged into a candidate halo were above the temperature threshold.

\begin{figure}
\begin{center}
\includegraphics[width=1\columnwidth]{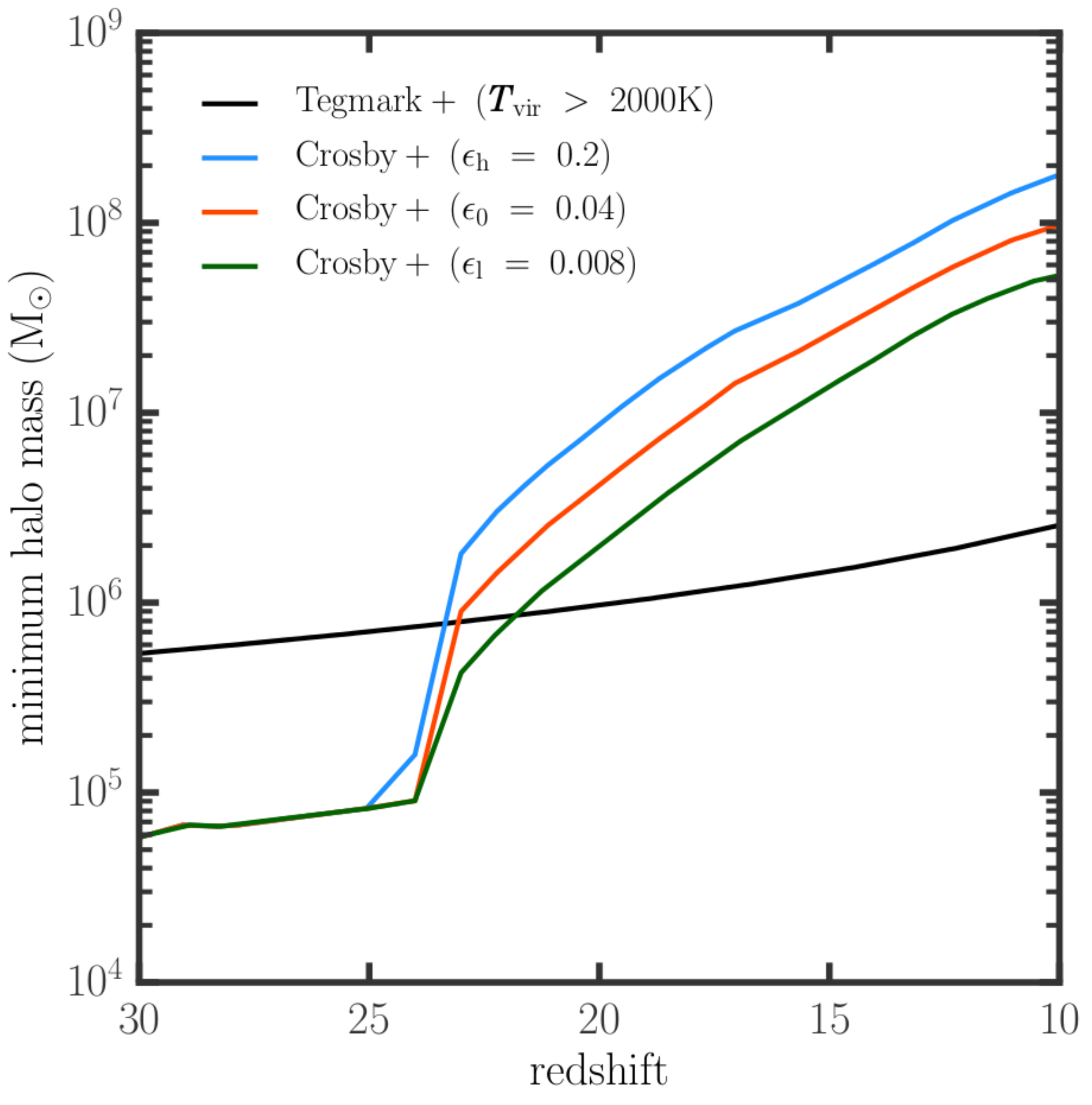}
\caption{{Minimum halo mass required for Pop\,III star-formation to proceed. We adopt two minimum mass thresholds for minihalo formation, one of which contains three variations of same semi-analytic model. Our first model is based on Tegmark et al. (1997). It requires that the H$_2$ cooling time is less than a Hubble time. For our second model, we interpolate the semi-analytic model of Crosby et al. (2013) which includes LW radiation produced by the first generations of stars in nearby halos at $z>20$. This results in a raising of the minimum mass depending on the  initial mass function adopted. In all models, progenitor halos are also checked to ensure that a candidate halo identified is the first in its history to go above the cooling threshold. The increase in the minimum mass at $z$ = 24 is due to the onset of Pop\,II star formation within the Crosby et al. (2013) model.
\label{fig:minimum_mass}
}}
\end{center}
\end{figure}

A critical feature required of a simulations attempting to identify minihalo candidates is the time between each snapshot used by the halo finder. To estimate whether we might be underestimating the number of candidate halos we compare the free-fall time of gas to our temporal resolution. We estimate that the free fall time of gas is $t_{ff}$ $\sim$ 0.1 * H(z), which for $z$ = 25 is $\sim$20 Myr, and $z$ = 10 is $\sim$70 Myr. Since our temporal resolution is $\sim$5 Myr between each snapshot, we are not under-counting any halos but we may be $over$-counting. By comparison, $Aquarius$ has a temporal resolution at these early times of $\sim$100 Myr which means Gao et al. may have under-counted the number of minihalos forming. If we assume $\sim$50 Myr is approximately the collapse time, and then compare this to the outer panel of Figure \ref{fig:mergetimescales}, we find $\sim$5$\%$ of minihalos merge within 50 Myrs, and $\sim$10$\%$ within 100 Myrs, meaning that we could be over-counting by $\sim$5$\%$, and work using the $Aquarius$ simulation will have undercounted by 5$\%$. 

\subsection{LW Feedback}

The minimum mass for collapse will be boosted to higher masses with the onset of LW radiation from Pop\,III stars which will photo-dissociate H$_2$ via the reaction H$_2$ + $\gamma_{\mathrm{LW}}$ $\rightarrow$ H + H, where $\gamma_{\mathrm{LW}}$ is a photon in the LW band of 11.12 -- 13.6 eV.  We must include this form of feedback in our model if we are to reliably determine which minihalo candidates represent the actual \textit{star-forming} halos at $z$ $>$ 10. Accordingly, we model the influence of a LW background via the semi-analytic model constructed by \citet{Crosby:2013bp}. We \textit{do not} explicitly calculate the relevant LW flux for each halo's stellar population, but simply adopt the adjusted minimum mass threshold for forming Pop\,III stars after including LW feedback. The \cite{Crosby:2013bp} model was based on simulations carried out using {\sc{enzo}}, an adaptive mesh refinement + $N$-body code. We have not carried out any simulation specific to our simulated volume but interpolate the minimum mass threshold they determined. Within their model, they followed 10 chemical species (H, C, N, O, Mg, Ca, Ti, Fe, Co, and Zn) in both the stellar and interstellar medium (ISM) components of every halo. The ISM was treated as a multiphase gas with a central region of dense, cold gas that is capable of forming stars and a hot, diffuse region exterior to the star-forming central region that is incapable of forming stars. For more details see the work of \cite{Crosby:2013bp,Crosby:2013ug}.

In Figure \ref{fig:minimum_mass}, we show the minimum host halo virial mass required as determined by \citet{Tegmark:1997ce} in order to cool to its virial temperature via H$_2$ cooling in the local Hubble time in \citet{Crosby:2013bp} then adopt three star formation efficiencies of $\epsilon$ = 0.008, 0.04, 0.2 (hereafter $\epsilon_{l}$, $\epsilon_{0}$, $\epsilon_{h}$) which adjust the minimum mass thresholds for collapse due to differing quantities of LW flux (J$_{21}$). \citet{Crosby:2013bp} adopted three different IMFs but since the star formation efficiency drives the Lyman-Werner flux over any particular selection of IMF we adopt parameterizations of three of their models distinguished by their star formation efficiencies only. The minimum mass thresholds we adopt for identifying Pop\,III star forming regions after including LW feedback are shown in Figure~\ref{fig:minimum_mass} (identical to Fig.~6 in \citealt{Crosby:2013bp}). The increase in the minimum mass threshold is particularly pronounced at z $\sim$ 24 where the onset of Pop\,II star formation from chemically enriched gas makes Pop\,II stars the dominant component of the stellar mass.

\subsection{Population\,II Star Formation}

In this section we describe criteria used for the formation of Pop\,II stars, which we assume form in the first galaxies.Once the virial temperature of the halo is high enough, atomic line cooling becomes important (\Tvir $\sim$ 10$^4$\,K). These halos are likely the sites of the first galaxies \citep{Bromm:2011dh} and as such we refer to all ``atomic cooling halos'' (ACHs) as first galaxies and vice-versa. The gas inflow rate into these systems largely traces the rate of inflow of the dark matter accretion rate, but this can be suppressed in the presence of an ionizing background. We adopt a simple model of reionization  following \cite{Bullock:2005is} whereby we divide atomic cooling halos into three populations based on their maximum circular velocity at the redshift set for reionization (assumed to be instantaneous at \zre = 10); (1) ACHs with \Vmax(z=10) $>$ 50\,\kms are not suppressed, (2) ACHs with 30\,\kms $<$ \Vmax(z=10) $<$ 50\,\kms are partially suppressed (i.e.,\ not all of their cold gas is star forming) and (3) ACHs with \Vmax(z=10) $<$ 30\,\kms are completely suppressed \citep{Thoul:1996gt}.

\subsection{Simple Chemical Enrichment Model}
\label{sec:chem}

After the accretion and collapse of cool gas in the central reservoir of a conducive halo, star formation proceeds, with the mass of each star set by the initial mass function (IMF). Some high-mass stars will eventually produce extremely energetic events such as pair-instability supernovae (PISN), whereby originally bound gas can be nearly entirely ejected (e.g. \citealt{Whalen:2004eq, Whalen:2008hu, Kitayama:2005ea}). If any of the proto-Milky Way's star forming regions were extremely clustered, this ejecta could likely pollute neighbouring halos and result in enhanced metal-line cooling spurring on subsequent star formation (e.g. \citealt{Smith:2015dw}). Detailed modelling of metal-enrichment of the subsequently formed first galaxies in these clustered environments shows that they can become significantly enriched to average metallicities of Z $>$ 10$^{-3}$ Z$_\odot$ \citep{Greif:2010du, SafranekShrader:2014io}. This inhomogeneous process can result in large spreads in chemical abundances of two to three orders of magnitude across the host system \citep{Wise:2008bq}. 

By definition, the first stars form out of chemically pristine gas. However, supernovae spew metals into the IGM \citep[e.g.][]{Madau:2001jj,Greif:2007ha,Jeon:2014iy,Smith:2015dw,OShea:2015fz}, and in some cases they can contaminate nearby minihalos that would otherwise be pristine \citep[e.g.][]{Smith:2015dw} (probably others). This separates minihalos into {\it endogenous} and {\it exogenous} minihalos, i.e. those that are initially unaffected by supernovae, and those that are externally enriched. In principle, this effect reduces the number of minihalos that should be considered as sites of Pop\,III star formation.

\begin{figure}
\begin{center}
\includegraphics[width=0.98\columnwidth]{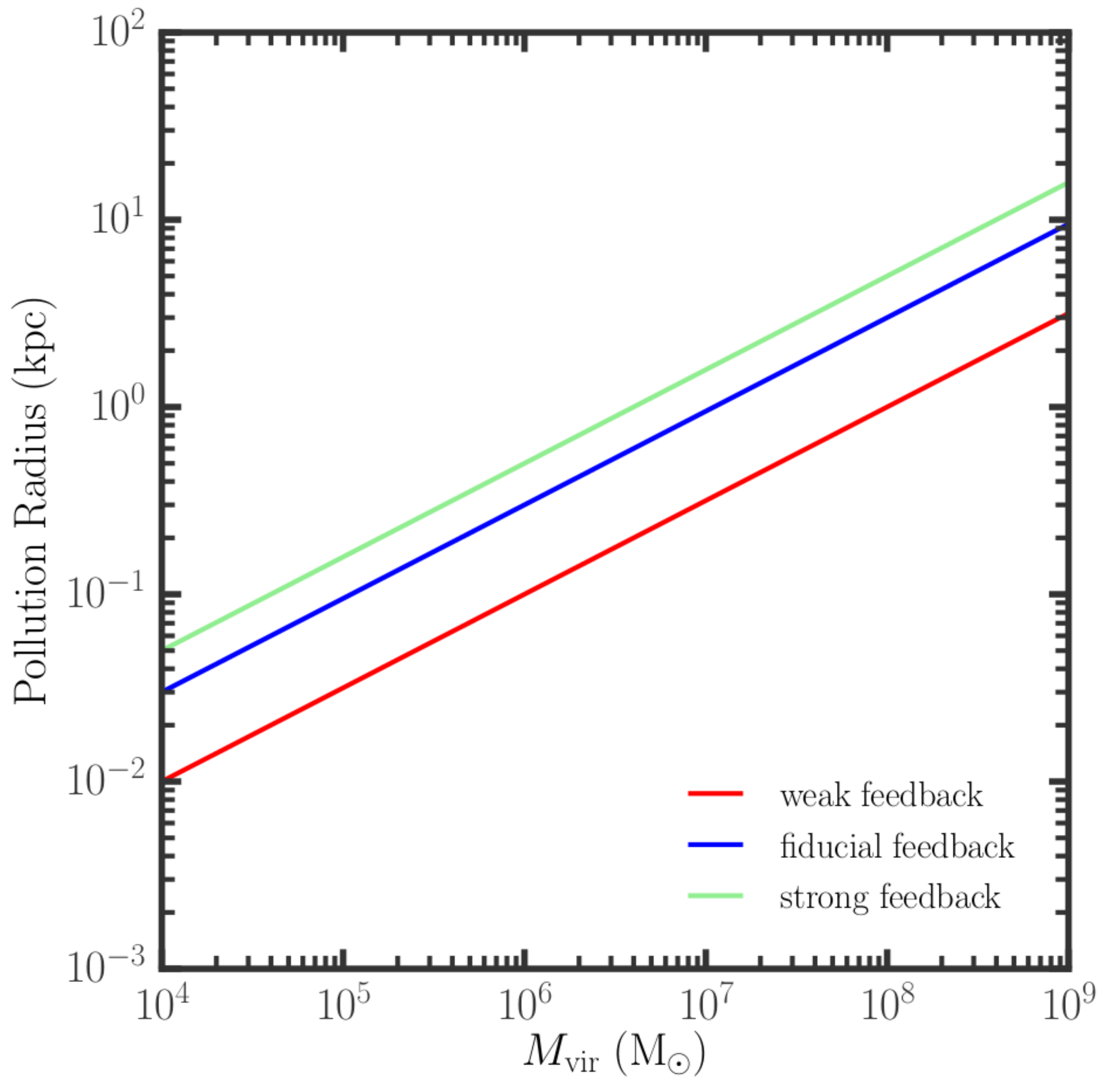}
\caption{{Simple chemical enrichment models with varying feedback. Our fiducial model yields a 300\,pc (physical) enrichment radius for 10$^5$\ \Msol halos \citep{Greif:2007ha,Ritter:2012bv,Ritter:2015dq,Smith:2015dw} and a 3\,kpc radius for 10$^8$\ \Msol halos \citep{Madau:2001jj}. We also adjust our normalization to account for strong feedback cases and weak feedback cases (e.g.\ for a 10$^6$\ \Msol halo the enrichment radius varies between 100\,pc and 500\,pc).
\label{fig:pollution}%
}}
\end{center}
\end{figure}

A complete characterization of metal pollution requires a fully hydrodynamic system \citep[e.g.][]{Greif:2010du,Wise:2012uc,OShea:2015fz}, but we can estimate the effect with a simple model based on distances between our halos. We consider a minihalo in our simulation to be exogenous (i.e., polluted) if its center is within the {\it pollution radius} of any other halo. For minihalos, the pollution radius is the size of a supernova remnant, which we take to be 300\,pc for a 10$^6$ \Msol halo \citep{Greif:2007ha, Ritter:2015dq, Smith:2015dw}. For an atomic cooling halo, the pollution radius is set based on the superbubble created by multiple supernovae associated with extended star formation, which we take to be 3\,kpc physical \citep{Madau:2001jj}. We then assume a mass-dependent pollution radius by taking the power law between these two points:

\begin{equation}
R_p = \frac{R}{R_8}\frac{M_{\mathrm{vir}}}{M_8}^\alpha,
\end{equation}
where $R_8$ is the pollution radius for a 10$^8$ \Msol halo (set to be 3\,kpc for the fiducial model), $M_8$ is 10$^8$ \Msol and $\alpha$ is the slope set by the $10^6$ \Msol halo pollution radius. Figure~\ref{fig:pollution} illustrates our fiducial model and two alternative normalizations allowing for stronger and weaker feedback. 

For simplicity, the pollution radii are assumed to be spherical, instantaneously grow to their maximum size, and instantaneously mix into any matter they encounter. However, detailed hydrodynamic runs find the metal enrichment is inhomogeneous and episodic \citep{Greif:2007ha, Ritter:2015dq, Smith:2015dw}, so we expect the number of externally enriched halos is an upper limit. It must be emphasized that we do not expect this simple enrichment prescription to accurately reflect the actual enrichment process of the first stars but to simply provide a broad-stroke model for gaining an understanding of the clustering properties and frequency of externally enriched objects.

\section{Results}
\label{sec:results}

\subsection{Visual Impression}

\begin{figure*}
\begin{center}
\includegraphics[width=0.85\textwidth]{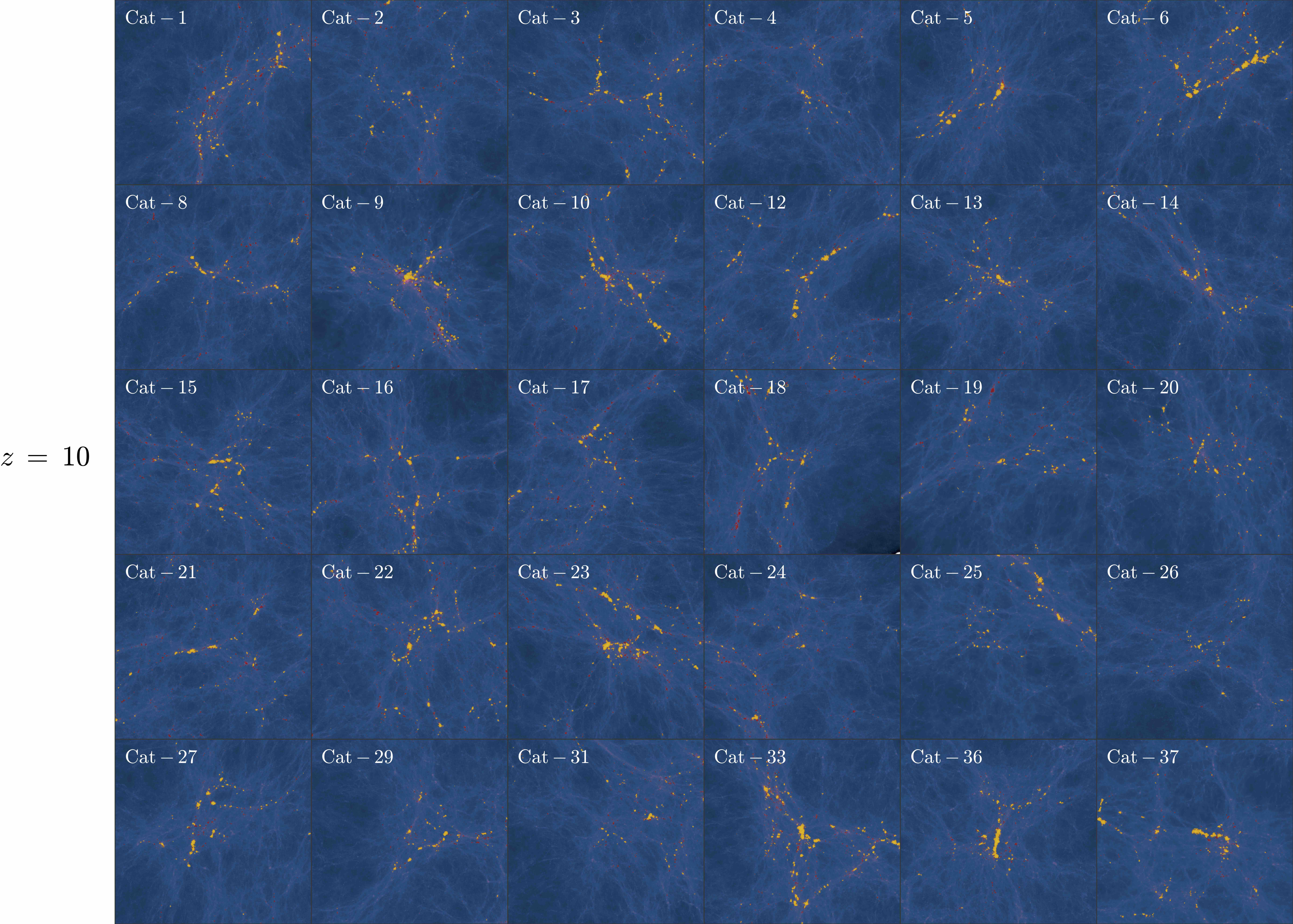}\\

\includegraphics[width=0.85\textwidth]{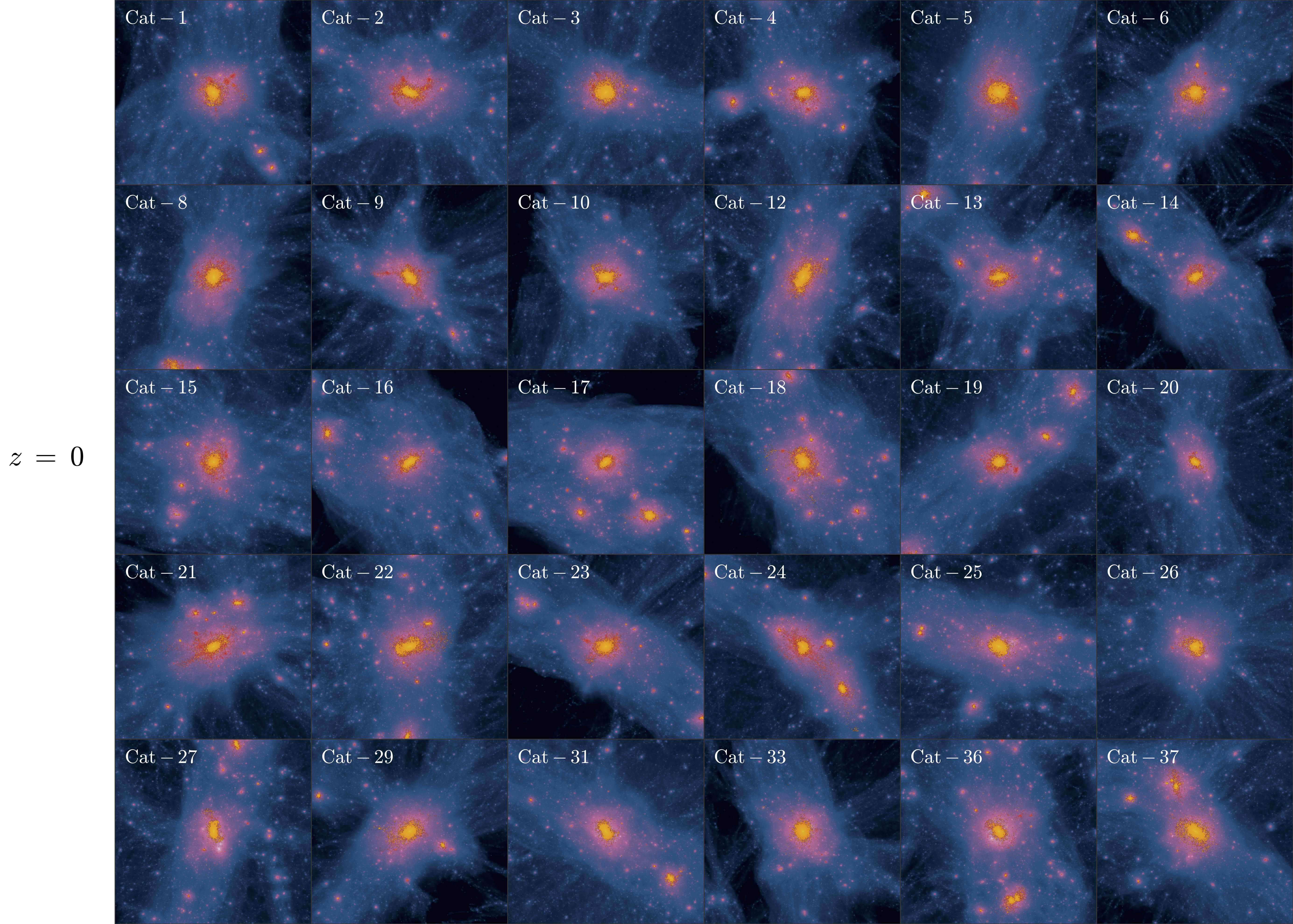}
\caption{{The 30 \caterpillar halos used in this study illustrate how the underlying dark matter distribution is overlaid with star particles. Particles are tagged as having formed within atomic cooling halos (red) and within molecular line cooling halos using our fiducial star formation efficiency ($\epsilon_0$ = 0.04). Five percent of the most bound particles were tagged for each respective system at formation. This is done purely for visualization purposes. Only halos which satisfy the temperature threshold before $z$ = 10 are tagged, as reionization is assumed to suppress star formation at $z$$ < $10. The top panels shows objects tagged at $z$ = 10 and the bottom panels are the same particles at $z$ = 0. The width of the image is 3 physical Mpc.
\label{fig:viz}%
}}
\end{center}
\end{figure*}

In Figure \ref{fig:viz}, we show the distribution of molecular cooling and atomic cooling halos across our sample of 30 \caterpillar simulation halos. The top five rows shows the distribution of these systems at $z$ = 10. We tag the 5$\%$ most-bound particles at formation, with H$_2$ cooling halos (``minihalos'') in yellow and atomic cooling halos in red. In this figure, we use the  LW feedback model adopting a star formation efficiency of $\epsilon_0$ = 0.04 to identify minihalos (Figure~\ref{fig:minimum_mass}).

The bottom five rows show the same respective particles at $z$ = 0 (image width is 1 physical Mpc in both cases). Halos are only tagged if they form before $z$ = 10 as we assume reionization suppresses star formation in all systems at these mass scales. Although, there are stark commonalities between halos at $z$ = 0, there are a wide variety of Lagrangian geometries at $z$ = 10. Some realizations at $z$ = 10 (e.g.\ Cat-2, Cat 9, Cat-36) show high densities of potentially star forming halos whilst other realizations show much more diffuse volumes of potentially star forming halos (e.g.\ Cat-1, Cat-6, Cat-33). In all cases, satellite systems both inside and outside the virial radius of the host contain potentially ancient stellar systems from the $z$ $>$ 10 era.

\subsection{Progenitors Of The Milky Way}

\subsubsection{Minihalo progenitors of the Milky Way}

\begin{figure}
\begin{center}
\includegraphics[width=0.98\columnwidth]{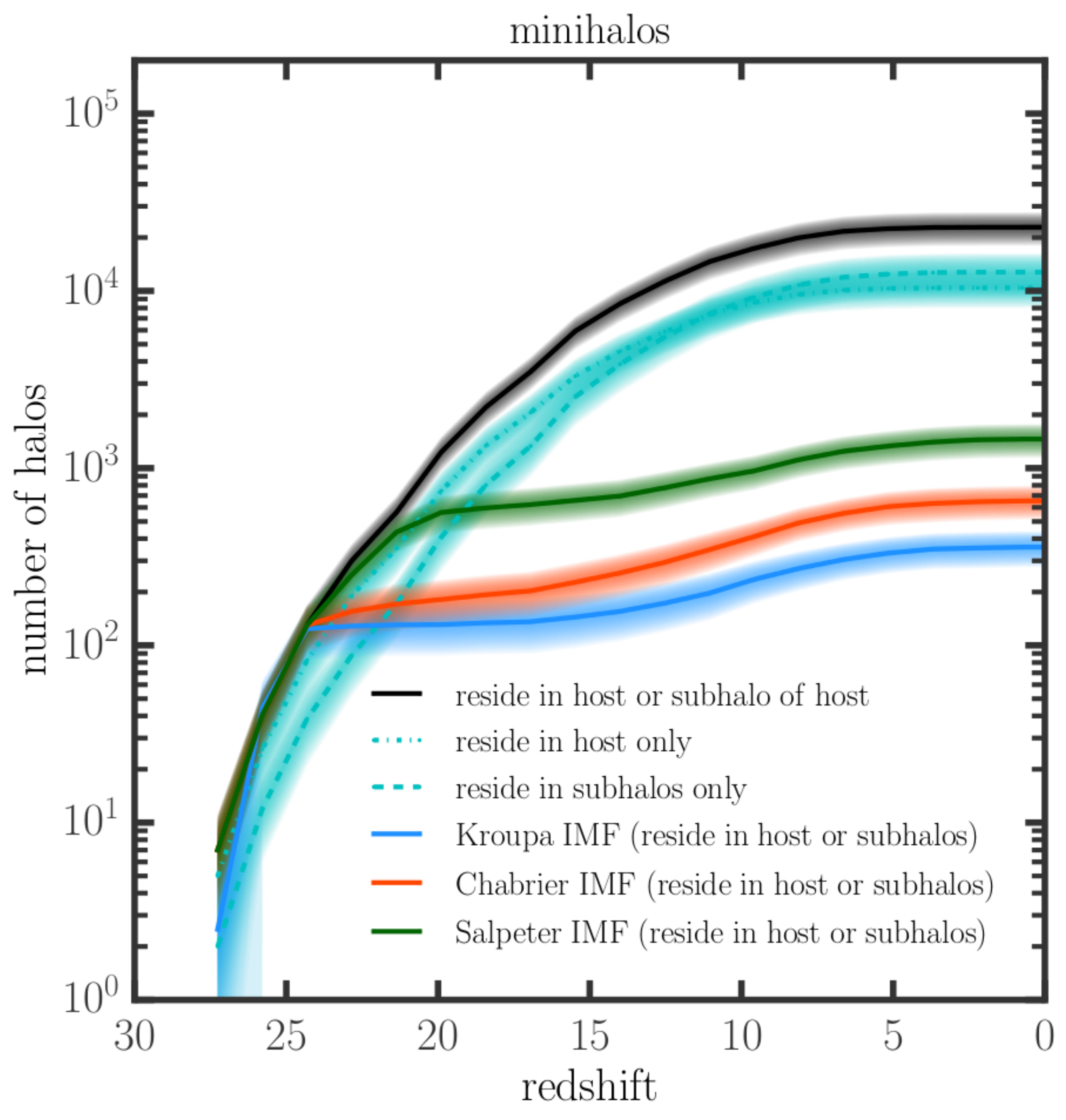}
\caption{{The cumulative number of Pop\,III star formation sites (``minihalos'') as a function of time averaged over all 30 \caterpillar halos. The onset of the second generation of star formation has a dramatic impact on total minihalo numbers as early as $z\sim25$, reducing the total number of potential star forming sites by 99.9$\%$, due to LW feedback. Over 90$\%$ of all minihalo sites have formed by $z$ = 10.
 \label{fig:numbersof_mh}}}
\end{center}
\end{figure}

In Figure \ref{fig:numbersof_mh}, we plot the cumulative number of minihalos formed over time. We only count the total number of systems which are \textit{accreted} into the central massive host and not those that end up residing in isolated halos at large galactocentric distances from the central host. The first of these Pop\,III star forming minihalos are identified at $z$ $\sim$ 26 and grow in number to approximately $\sim$23,000 total (black line) potential sites assuming the \cite{Tegmark:1997ce} temperature minimum mass criteria (\Tvir $\sim$ 2000\,K). The shaded regions for each line indicate 1-$\sigma$ across all 30 halos in our sample for each of the methods of identification. There is $\sim$20$\%$ scatter in the total number at nearly all times but some can be attributed to the fact that larger mass hosts have more progenitors on average ($n/10^{12}$ \Msol  $=\ 1.08\ \times\ 10^{-8}\ \pm\ 0.03\ \times\ 10^{-8}$, where $n$ is the total number of systems). 

Further dividing this population into progenitor systems that ultimately end up in the main host halo or in any of the subhalos of the main halo, we find that at $z=0$ roughly the same number of progenitor minihalos end up in the central host (45 $\pm$ 11$\%$, 10403 $\pm$ 2418) as in the host's subhalos (55 $\pm$ 16$\%$, 12746 $\pm$ 3568).

Although there are a large number of potential Pop\,III star forming sites, the first luminous ones to have formed will greatly impact candidate sites for subsequent star formation due to the onset of the LW background. In Figure \ref{fig:numbersof_mh}, we also show the cumulative number of halos which could have still collapsed in the presence of this LW background. Table \ref{tbl:number_mhs} shows the cumulative number of halos for each population identified. We find drastic reductions by as much as 98$\%$ of potential star forming sites which would have otherwise cooled and collapsed via molecular line cooling in the absence of a LW background. Altering the choice of the star formation efficiency, $\epsilon$, changes the amount of LW flux and consequently the potential number of sites from $\sim$358 $\pm$ 82 (1-$\sigma$, $\epsilon_h$) to $\sim$1458 $\pm$ 314 (1-$\sigma$, $\epsilon_l$). Between all cases, a minimum of 94$\%$ of the potential number of halos, which are nevertheless later accreted into the central host, are prevented from forming stars. For all three star formation efficiencies, approximately $\sim$50$\%$ end up in subhalos and $\sim$50$\%$ end up in the primary host by $z$ = 0.

\begin{table}
\caption{{Number of minihalos across all of the \caterpillar halos, broken down by final location at $z$ = 0 and by the additional use of different star formation efficiencies (including $\pm$1-$\sigma$ variance).}}

\begin{center}
\begin{tabular}{lcc}
\hline
Selection & Number & Fraction \\
\hline                      
Reside within host or subhalos	 & 22856 $\pm$  4915 & 1.00 $\pm$ 0.22
\\
Reside in host 				 & 10403 $\pm$  2418 & 0.45 $\pm$  0.11\\
Reside in subhalos 			& 12746 $\pm$  3568 & 0.55 $\pm$ 0.16 \\
\hline
incl. LW ($\epsilon_h$ = 0.2) 	 & 358 $\pm$    82 & 0.02 $\pm$ 0.00\\
incl. LW ($\epsilon_0$ = 0.04)   & 653 $\pm$   141 & 0.03 $\pm$ 0.01\\
incl. LW ($\epsilon_l$ = 0.008)	 & 1458 $\pm$   314 & 0.06 $\pm$ 0.01\\
\hline
\end{tabular}
\end{center}
\label{tbl:number_mhs}
\end{table}

\subsubsection{Atomic cooling halo progenitors of the Milky Way}

\begin{figure}
\begin{center}
\includegraphics[width=0.98\columnwidth]{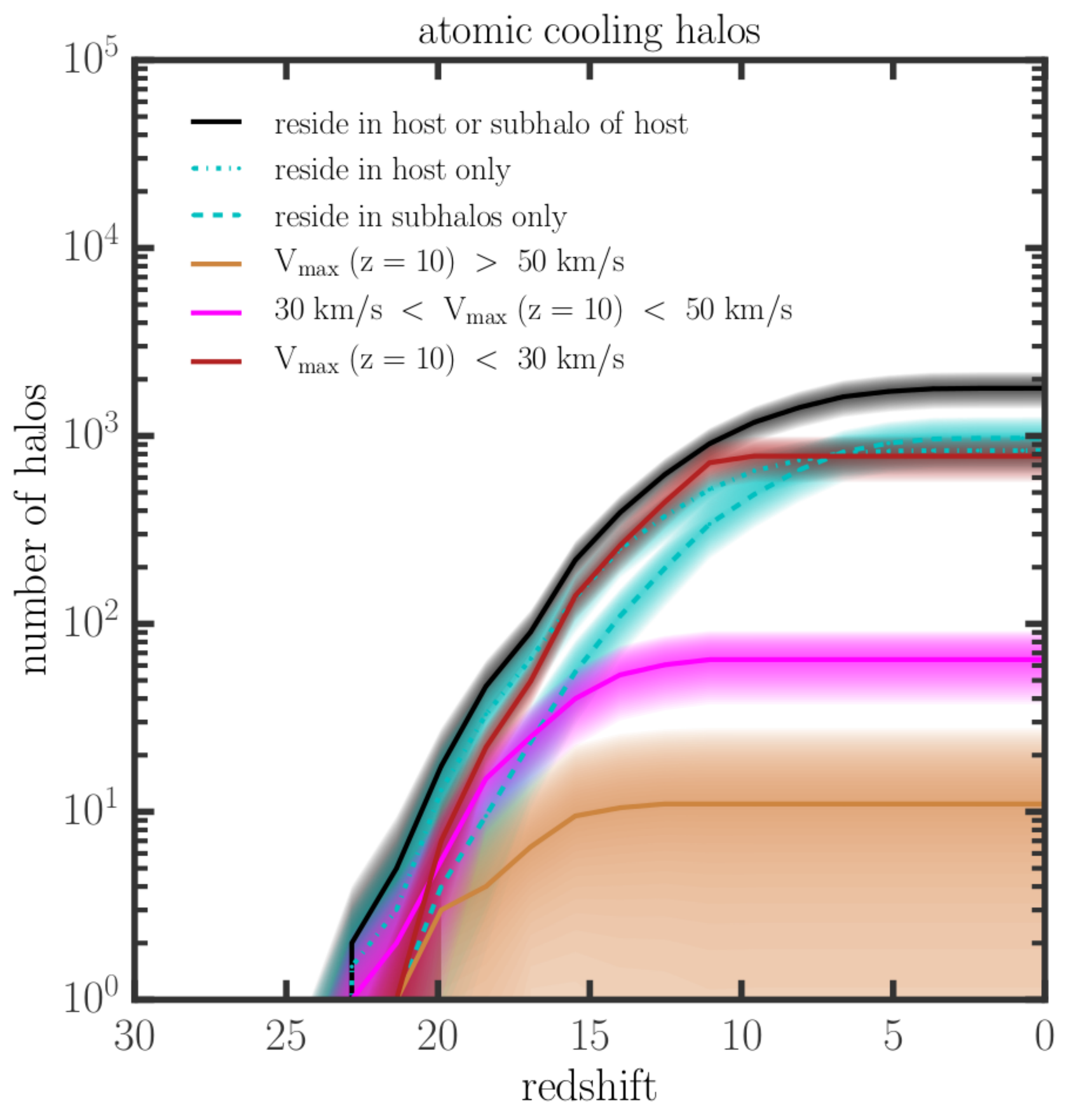}
\caption{{Number of atomic cooling halos which reside in either subhalos or halos by $z$ = 0. We further divide the population into those which are suppressed, partially suppressed or fully star forming based on their maximum circular velocity of their descendants at $z$ = 10. On average, 781 $\pm$ 214 potential atomic cooling halos are suppressed and stop forming stars due to the reionization background. Approximately $\sim$11 survive the reionization era and will continue to form stars provided there exists a supply of cold gas. We find 64 $\pm$ 27 are partially suppressed and will only convert some fraction of their cold gas into stars. Just over half of all atomic cooling halos to have formed reside within the central host (54\% $\pm$ 16\%) in the present day whilst the remainder (46\% $\pm$ 11\%) reside in subhalos).
\label{fig:numbersof_fg}%
}}
\end{center}
\end{figure}

In Figure~\ref{fig:numbersof_fg}, we plot the total number of halos which satisfy the virial temperature condition (\Tvir $>$ $10^4$\,K). We divide the population into five categories, three of which are a subset of just one. We only count atomic cooling halos which end up in the central host or in a subhalo of the central host by $z=0$. Of the subset that accretes into the primary host and subhalos, we further divide them into three groups; (1) halos with \Vmax($z$ = 10) $>$ 50\,\kms are not suppressed (green), (2) ACHs with 30\,\kms $<$ \Vmax($z$ = 10) $<$ 50\,\kms are partially suppressed (blue), and (3) halos with \Vmax($z$ = 10) $<$ 30\,\kms are completely suppressed (red).

Table~\ref{tbl:numbersof_fg} lists the cumulative number of halos which form in each category. We find that approximately 1793$\pm$396 (1-$\sigma$) halos within a Milky Way sized system satisfy the atomic cooling limit and are eventually accreted either into the host itself or its subhalos. As with the minihalos, we find that approximately half (45 $\pm$ 11$\%$) reside within the central host and half (55 $\pm$ 16$\%$) reside within subhalos at the present day. Nearly half the halos that surpass the atomic cooling limit for the first time in their main branch end up within halos below the suppression scale at $z$ = 10. We find approximately 11 halos (per host) with $V_{\mathrm{max}}$($z$ = 10) $>$\,50\,\kms at $z$ = 10 that will continue to form stars provided there exists a supply of cold gas. Some of these will merge with other halos before being accreted by the central host. These halos, which are not suppressed, combined with any of the partially suppressed ones in the post reionization era (64 $\pm$ 27 that will only convert some fraction of their cold gas into stars), could go on to become present day dwarf spheroidal galaxies around the Milky Way.

\begin{table}

\caption{{Number of atomic cooling halos across all of the \caterpillar halos broken down by the various models for identification at $z$ = 0 (including $\pm$1-$\sigma$ variance).}}
\begin{center}
\begin{tabular}{lcc}
\hline
Selection & Number & Fraction\\
\hline                            
Reside in host or subhalos				& 1793 $\pm$  396 & 1.00 $\pm$ 0.22\\
Reside in host 				& 973 $\pm$  290 & 0.54 $\pm$ 0.16\\
Reside in subhalos 				& 836 $\pm$  206 & 0.47 $\pm$ 0.11 \\
\hline
No suppression$^a$  		& 11 $\pm$   16 & 0.01 $\pm$ 0.01\\
Partially suppressed$^b$ 	& 64 $\pm$   27 & 0.04 $\pm$ 0.02\\
Fully suppressed$^c$ 		& 781 $\pm$  214 & 0.44 $\pm$ 0.12\\
\hline
\end{tabular}\\
$a$:\ $V_{\mathrm{max}}$($z$ = 10) $>$ 50\,\kms.\\
$b$:\ 30 $<$ $V_{\mathrm{max}}$($z$ = 10) $\leq$ 50\,\kms.\\
$c$:\ $V_{\mathrm{max}}$($z$ = 10) $\leq$ 30\,\kms. A total of 937 halos per host form after $z$ = 10 have \Vmax $<$ 30 km/s and are assumed to be suppressed.
\end{center}
\label{tbl:numbersof_fg}
\end{table}%

\subsection{When were the first stellar systems accreted into the Milky Way?}

\begin{figure}
\begin{center}
\includegraphics[width=0.98\columnwidth]{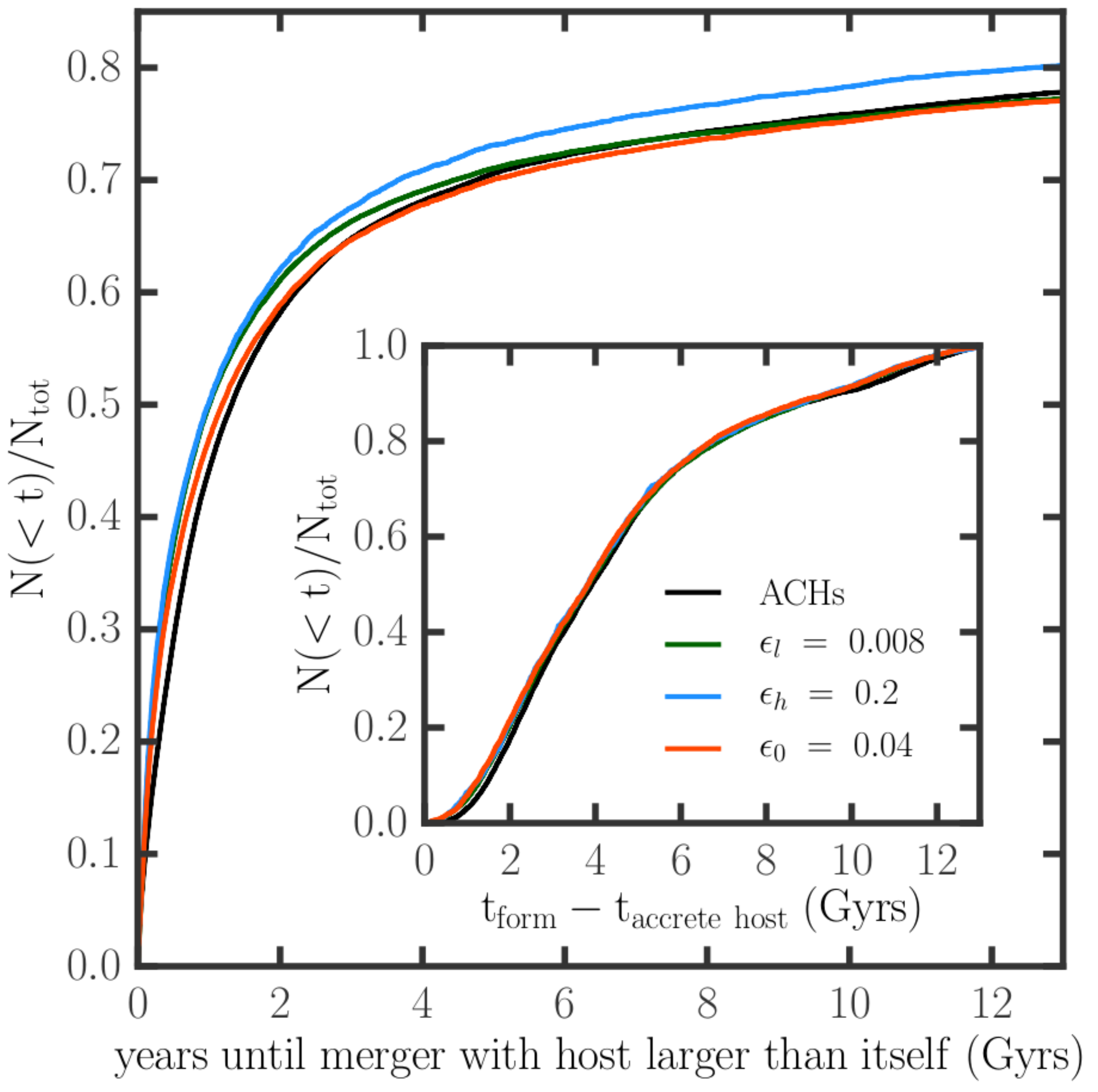}

\caption{{\textit{Outer panel:} The cumulative distribution function of the time of the first merger of all atomic cooling halos and minihalos. Approximately 20$\%$ of all identified halos do not merge with any other halo larger than itself other than merging with the main host. \textit{Inner panel:} The cumulative distribution function of the time between when halos form and when they enter the virial radius of the central host but have not merged by $z$ = 0. Only $\sim$50$\%$ of halos enter the host's virial radius within 4 Gyrs, indicating that many systems evolve in isolation for a significant portion of their lifespan.
\label{fig:mergetimescales}%
}}
\end{center}
\end{figure}

In Figure~\ref{fig:mergetimescales}, we show the cumulative distribution function at the time of the first merger for all identified systems which end up within the virial radius of the host at $z$ = 0 (across all \caterpillar halos). Approximately 50$\%$ of minihalos and atomic cooling halos merge into another halo larger than itself within 1 Gyr. Approximately 22 $\pm$ 1 $\%$ of all atomic cooling halos never have a merger with another halo larger than itself along its main branch. Similarly, approximately 20$\%$ of all minihalos never merge with another host larger than itself along its main branch. For the LW model adopting a high star formation efficiency ($\epsilon_h$) this fell to 19 $\pm$ 3 $\%$. For the low star formation efficiency ($\epsilon_l$) and fiducial star formation efficiency ($\epsilon_0$), they both yielded 22 $\pm$ 3 $\%$. 

In the inner panel of Figure~\ref{fig:mergetimescales}, we also show the time between formation and accretion for objects which do not merge with anything larger than itself, i.e.\ the subset of halos amounting to $\sim$20$\%$ of halos that have not merged in the outer panel. We find that 50$\%$ of all atomic cooling halos and minihalos accrete in the host within 4\ Gyrs and 80$\%$ are accreted within 8\ Gyrs. When compared to the history of an average subhalo at $z$ = 0, atomic cooling halos and minihalos systematically cross the virial radius of the central host at earlier times as they were the first halos to form.

\subsection{Spatial Distribution $\&$ Clustering}
\label{sec:clustering}

\begin{figure}
\begin{center}
\includegraphics[width=0.98\columnwidth]{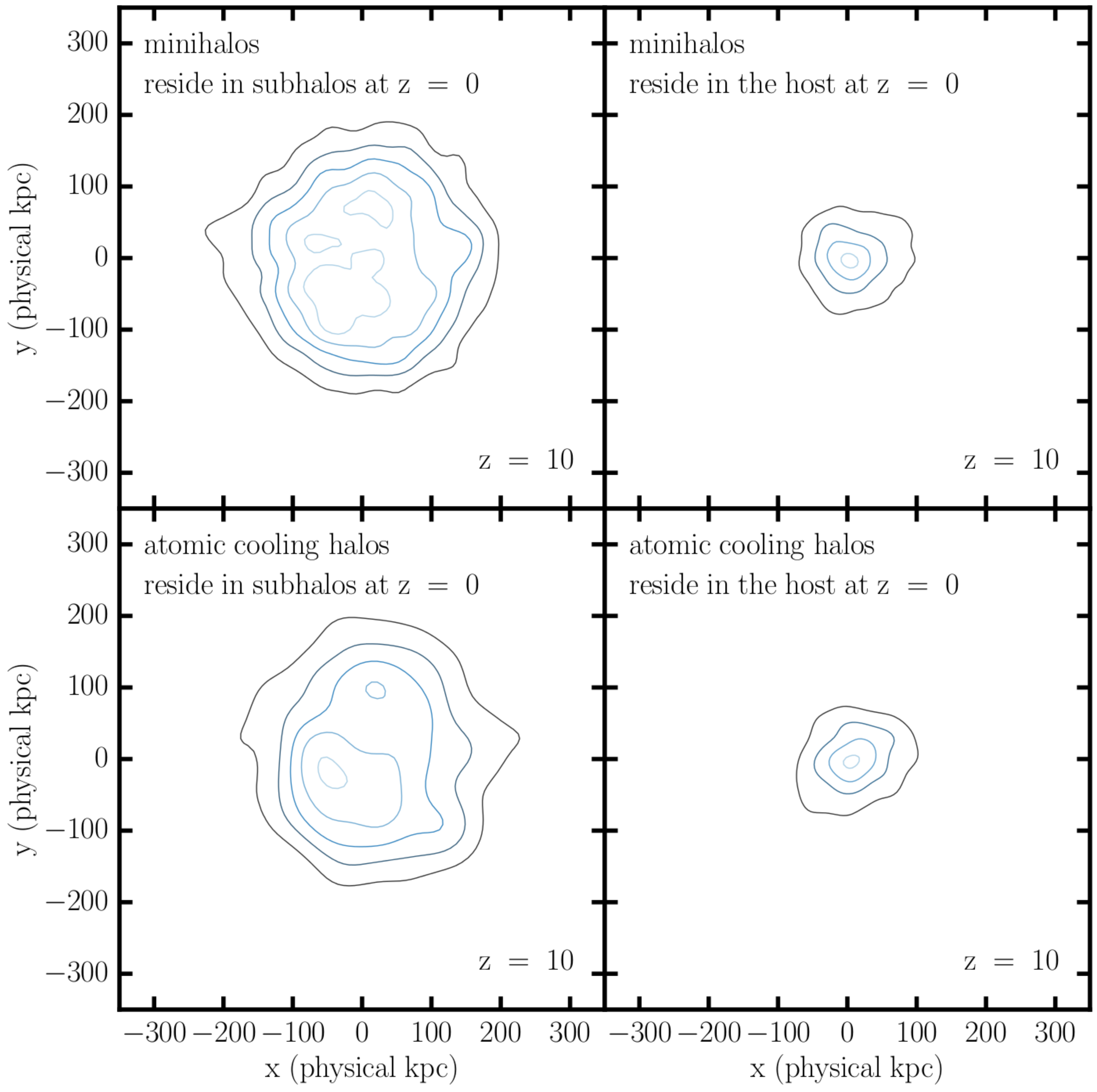}
\caption{{Contour plot of the positions of all minihalos and atomic cooling halos relative to the host (physical distance) identified in a single snapshot at $z$ = 10. The top rows represent the positions of minihalos while the bottom rows represents the atomic cooling halos. The first column represents systems that reside in subhalos at the present day while the right column represents systems that reside in the central host at the present day. This characteristic spatial correlation between present day environment and formation environment is clear for all times -- the initial stellar systems residing in subhalos today were much less clustered at earlier times compared to their counterparts that end up in the central host. This diagram represents the stacked positions for all 30 \caterpillar halos in our sample.
\label{fig:kde}%
}}
\end{center}
\end{figure}

We investigate the spatial distribution of minihalo and first galaxy progenitors of Milky Way sized systems. In Figure~\ref{fig:kde}, we demonstrate the spatial clustering of objects which end up inside subhalos or the central host at $z$ = 0. This figure shows the density contours of all systems identified as minihalos (using the Tegmark et al. (1997) prescription) and atomic cooling halos found in a single snapshot corresponding to $z$ = 10 across all 30 of our \caterpillar halos (the spatial distributions for minihalos identified with LW feedback are the same, see Section~\ref{sec:caveats}). Across all \caterpillar halos, we find that objects whose descendants eventually reside within subhalos are much less compactly clustered at high redshift than their counterparts that ultimately end up in the central host.

\begin{figure*}
\begin{center}
\includegraphics[width=0.98\textwidth]{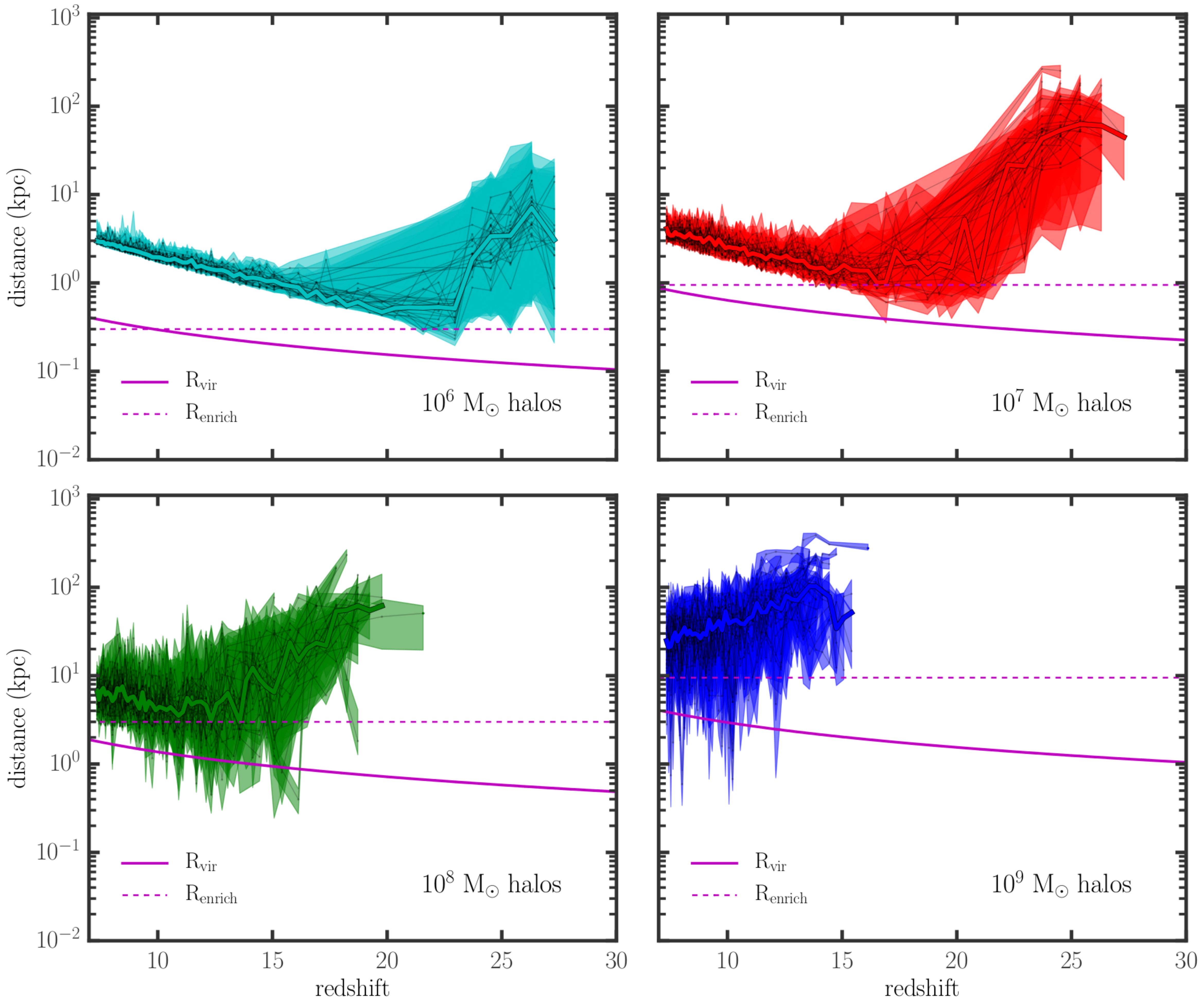}

\caption{{Median minimum distance (physical) of every identified minihalo to any different mass halos (star forming) as a function of time. Each minihalo was identified using the minimum mass threshold found by \citet{Crosby:2013bp} which includes LW feedback using our fiducial star formation efficiency ($\epsilon_0$ = 0.04). Each of the thin black lines represent a single \caterpillar simulation and the shaded region represents the 1-$\sigma$ variance. The solid line represents the median of all 30 $Caterpillar$ runs. In each panel, the cyan line underneath represents the virial radius of halos with 10$^{6-9}$ \Msol, based on the \citet{Bryan:1998cc} prescription. The dashed line is the enrichment radius for each of these halos calculated with our fiducial enrichment model. It is clear that the majority of the external enrichment of a given minihalo is driven by neighbouring larger mass halos, e.g.\ 10$^{8-9}$\ \Msol. Each of the neighbouring halos are checked to ensure they are actually star forming by determining if any progenitors contain accreted halos that have satisfied the virial temperature criterion. The median minimum distance is an indicator of the density of star forming halos. Starting at high-$z$, the density first increases due to a proliferation of galaxy formation. Later, the formation rate of new galaxies declines and the Hubble expansion begins to dominate, leading to a decrease in density (or increase in distance). This turnaround point occurs at higher $z$ for lower mass halos since low mass halos form earlier in the universe than high mass halos.
\label{fig:clustering}
}}
\end{center}
\end{figure*}

To get a better understanding of the separation properties of  minihalos to their neighbouring star forming halos, we plot in Figure~\ref{fig:clustering} how far away star forming halos are located from each of the minihalos. The halos are separated into increasing mass bins. In the first panel, we show the median minimum distance of each minihalo to all other star forming 10$^6$ \Msol halos. We find most of them are several virial radii away from any minihalo at all times. The median minimum distance to a 10$^6$ \Msol star forming halo at $z$$\sim$20 is $\sim 1$\,kpc (physical) indicating the proto-Milky Way formed in a very clustered environment. There is, however, large scatter in the median minimum separation ranging from 800\,pc to 3\,kpc at $z$ = 20 across each of the \caterpillar simulations. The larger neighbouring star forming halos (10$^{8-9}$ \Msol) often have minihalos residing within a few virial radii during the time of their formation. This often leads to minihalos experiencing external chemical enrichment coming from these neighboring halos during their initial fragmentation process. But details depend on the individual case since there is significant scatter of several kiloparsecs of the median minimum separation at $z$$\sim$20.

\subsection{Internally $\&$ Externally Enriched Fraction}

We have shown clearly that there is a spatial preference for progenitors of subhalos when compared to the progenitors of the central host in the high-redshift era of the Milky Way. These spatial biases are expected to manifest themselves in the chemical enrichment history of their respective stellar constituents as systems that reside in the host today come from more clustered environments. We apply our simple chemical enrichment model from Section~\ref{sec:chem} to determine what fraction of minihalo progenitors of the Milky Way were likely externally or internally enriched. These processes lead to two classes of systems in the proto-Milky Way era; \textit{endogenous} systems (chemically enriched solely by internal processes) and \textit{exogeneous} systems (enriched by internal \textit{and} external processes).

\begin{figure*}
\begin{center}
\includegraphics[width=0.95\textwidth]{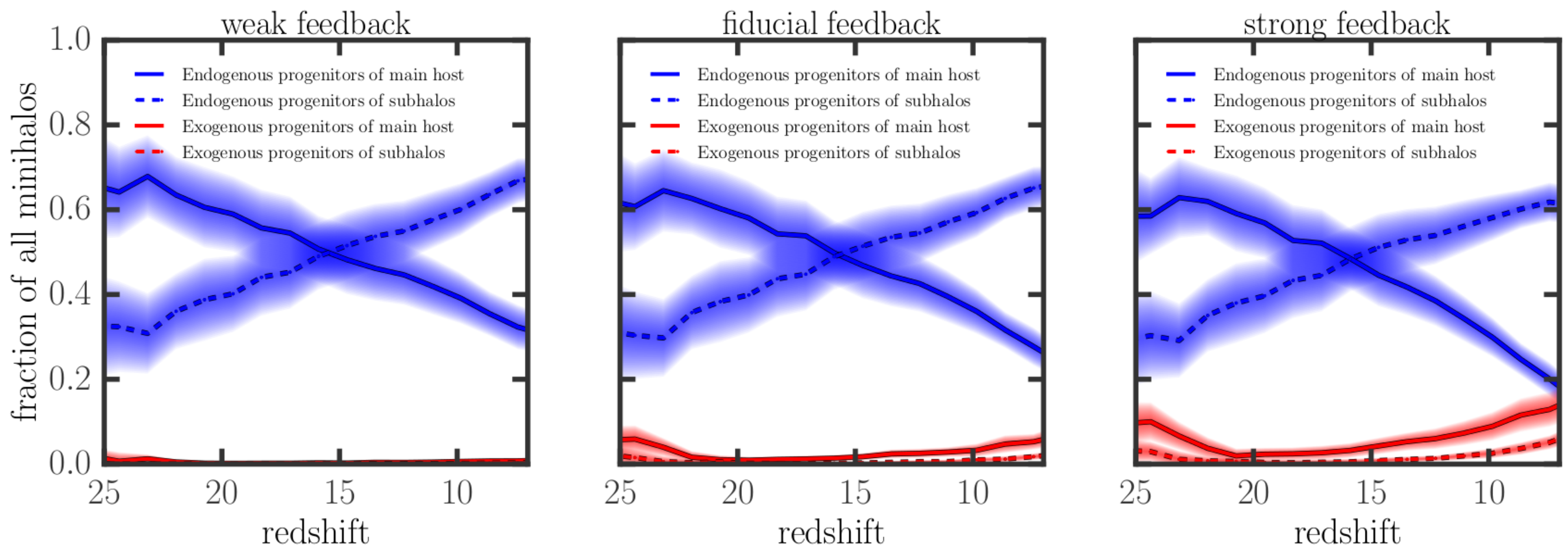}
\caption{{Fraction of exogenous (enriched by an external system) and endogenous (enriched by internal processes) progenitor minihalo systems for each of the Milky Way-mass halos in our sample. Each panel represents a different strength of feedback (see Figure~\ref{fig:pollution}). The total exogenous population varies from $\sim 1\%$ of all halos at $z=7$ in the weak feedback model to $18\%$ in the strong feedback model. Minihalos that form early, near $z=25$, are more likely to be progenitors of the host since they have more time to be accreted, lose angular momentum, and get disrupted, whereas minihalos forming later, near $z=7$ are more likely to be progenitors of subhalos since they don't have enough time to be fully disrupted in the host. 
\label{fig:enrichedfrac}%
}}
\end{center}
\end{figure*}

In Figure~\ref{fig:enrichedfrac}, we show what fraction of the total population are exogenous or endogenous as a function of time for the minihalos identified via the Tegmark et al. (1997) prescription. We further break this population down into minihalos which end up in subhalos and minihalos which end up in the main host. The breakdown of populations is similar in each of the \textit{Caterpillar} halos in the sample. The feedback prescription used in the middle panel is our fiducial model whereby halos with \Mvir = 10$^6$ \Msol have enrichment radii of 300\,pc while halos with \Mvir = 10$^8$ \Msol have enrichment radii of 3\,kpc. We also show results for the weak and strong feedback models from Figure~\ref{fig:pollution}.

At $z$ = 20, we find that an overwhelming proportion of the minihalo population are endogenous systems, evolving in isolation, for all three feedback models. This continues to later times and only by $z$ = 7 do we observe any significant number of minihalos becoming exogenous, or externally enriched. In the strong feedback model, the fraction of exogenous minihalos rises from just $3\%$ at $z=20$ to $18\%$ at $z=7$. Using the weak feedback model, merely $\leq 1\%$  of minihalos are externally enriched between $z=20$ and $z=7$. Meanwhile, the endogenous population flips from being dominated by progenitors of the host at $z=20$, to being dominated by progenitors of subhalos at $z=7$. This is due to a bias where halos that form earlier have more time to be pulled into the central host and disrupted by $z=0$ than halos which form later.

In Table~\ref{tbl:contamstat}, we list the breakdown of minihalos for the strong feedback model into raw percentages. When restricting the sample to just progenitors of the host, the fraction of exogenous halos begins approaching that of endogenous halos towards $z=7$. At $z=20$, $3\%$ of progenitors of the host are exogenous. By $z=7$, that number jumps to $40\%$. When restricting to progenitors of subhalos, the fraction of endogenous systems reaches a much smaller peak of $7\%$ at $z=7$. This is caused by the spatial biasing in the assembly history of the host. The progenitors of accreted systems which reside within the host in the present day tend to be more centrally clustered in the most over-dense regions, leading to a higher probability that the enrichment bubbles of nearby systems overlap with the surrounding halos. Furthermore, those systems which end up in the present day host are more likely to be externally enriched the later they form due to a combination of spatial clustering and a greater abundance of larger, $10^9$\Msol, star forming halos (see Figure~\ref{fig:clustering}).

\begin{table*}
\caption{{Fraction of halos which are exogenously or endogenously enriched for the fiducial feedback model at different times for 30 \caterpillar halos ($\pm$1-$\sigma$ variance).}}
\begin{center}
\begin{tabular}{ccccc}
\hline
Type & $z$ = 20  & $z$ = 15 & $z$ = 10  & $z$ = 7  \\
\hline
${\rm Endogenous\ progenitors\ of\ main\ host}$       & 0.57 $\pm$ 0.08 & 0.45 $\pm$ 0.07 & 0.30 $\pm$ 0.05 & 0.20 $\pm$ 0.04 \\
${\rm Endogenous\ progenitors\ of\ subhalos}$   & 0.40 $\pm$ 0.09 & 0.51 $\pm$ 0.08 & 0.58 $\pm$ 0.05 & 0.62 $\pm$ 0.05 \\
${\rm Exogenous\ progenitors\ of\ main\ host}$        & 0.02 $\pm$ 0.02 & 0.04 $\pm$ 0.02 & 0.09 $\pm$ 0.03 & 0.13 $\pm$ 0.04 \\
${\rm Exogenous\ progenitors\ of\ subhalos}$    & 0.00 $\pm$ 0.01 & 0.01 $\pm$ 0.01 & 0.03 $\pm$ 0.01 & 0.05 $\pm$ 0.02 \\
\hline
\end{tabular}
\end{center}
\label{tbl:contamstat}
\end{table*}%

\subsection{Remnants of the first stellar systems in dwarf galaxies}

The progenitor halos of the Milky Way can be split into two distinct populations: (1) ``halo progenitors'' (i.e.,\ those that formed, merged and accreted, subsequently dispersing throughout the stellar halo of the Milky Way) and (2) ``dwarf progenitors'' (i.e., those that formed, accreted and merged into what are now dwarf galaxies). The progenitor merger tree of each of these two systems will invariably be littered with minihalos and atomic cooling halos. 

This presents an opportunity to consider in detail the origin and nature of the observable dwarf satellite galaxies of the Milky Way. Especially their early chemical composition, and consequently also that of their oldest, most metal-poor stars must have been driven by the total number of high-$z$ minihalos and atomic cooling halos that each dwarf galaxy accreted throughout its evolution. 

\begin{figure*}
\begin{center}
\includegraphics[width=0.98\textwidth]{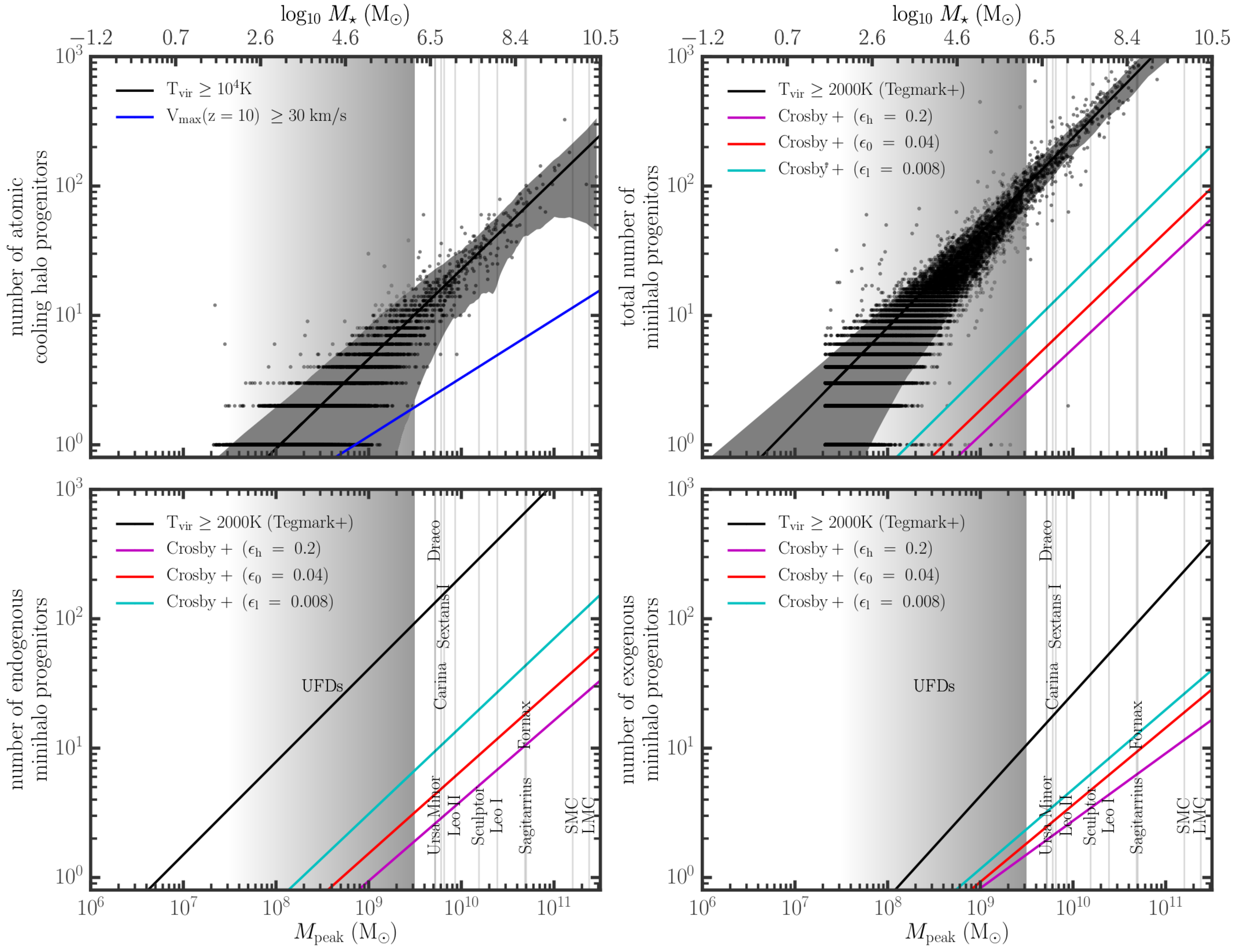}

\caption{{Total number of progenitors of a given subhalo (\textit{top left}: atomic cooling halos, \textit{top right}: total, \textit{bottom left}: endogenous minihalos, \textit{bottom right}: exogenous minihalos) as a function of peak subhalo mass. Typical 1-$\sigma$ variance for each fit (across all 30 $Caterpillar$ simulations used in this study) is shown in the top left/right panel (grey band). These are omitted for the other fits for the sake of clarity. The peak mass corresponds to a stellar mass as determined from the abundance matching prescription of \citet{GarrisonKimmel:2014bca}. As a guide, we have drawn vertical lines corresponding to the stellar mass of each of the observed nine classical dwarf spheroidal galaxies. Although more massive dwarf galaxies tend to have a large number of minihalo progenitors, the total number depends strongly on the inclusion of the LW feedback. This reduction is particularly pronounced for ultra-faint dwarf galaxies, with 90$\%$ fewer potential minihalo progenitors. There are slightly fewer ($\sim$10$\%$) endogenous progenitors (i.e.\ progenitors which have evolved in isolation) at fixed subhalo peak mass. The uncertainty in LW models is similar to the halo-to-halo scatter. As previously stated (e.g.\ Sawala et al. 2014), estimates of stellar mass based on abundance matching are unreliable for \Mpeak $\leq$ 10$^9$ \Msol. We only estimate the number of progenitors for UFDs (range highlighted in green) by extrapolation, which as such, is speculation.
\label{fig:progenitors}}}
\end{center}
\end{figure*}

Since our simulation suite runs until $z$ = 0, we can determine how many candidate minihalos and atomic cooling halos have merged with a given dwarf galaxy since its formation. In Figure~\ref{fig:progenitors}, we plot a parameterized fit to the number of progenitors of a given subhalo in the present day for all \caterpillar simulations. We relate the number of each respective system (i.e.,\ exogenous and endogenous systems for each definition of minihalo and atomic cooling halo) to the peak mass of present day subhalos via the following form,

\begin{equation}
n_p\ =\ n_0\left(M_{\mathrm{peak}}\right) ^\alpha,
\end{equation}

where $n_p$ is the number of progenitor systems and \Mpeak is the peak mass along the main branch of a given subhalo. $n_0$ is a normalization quantity. Given this functional form, our best fit estimates are presented in Table~\ref{tbl:param}. We also show  stellar mass estimates for these systems using the \citet{GarrisonKimmel:2014bca} abundance matching prescription as a reference. We find that the number of minihalo progenitors at a fixed \Mpeak (peak mass along the main branch) depends very much on whether LW feedback is included. Without LW feedback, a halo with a peak mass of 10$^9$ \Msol (\Mstar $\sim$ 10$^4$ \Msol) would have accreted $\sim$30 minihalo progenitors. With LW feedback, a halo with the same peak mass would actually only have accreted $\sim$10 halos at most. This is particularly pronounced at even lower peak masses (e.g.,\ UFDs), where one expects less than one minihalo to have been accreted into the system by $z$ = 0 when including LW feedback.

\begin{table*}
\begin{center}
\caption{{Number of progenitors for a given halo at $z=0$ ($\pm$1-$\sigma$ variance across halos).}}

\begin{tabular}{lcccc}
\hline
Number Of Progenitors & n$_0$ (\Msol) & $\pm$1-$\sigma$ & $\alpha$ & $\pm$1-$\sigma$ \\
\hline
\textbf{Minihalos} 			&  & &  \\
\citet{Tegmark:1997ce} (\Tvir $>$ 2000\,K) & 1.107$\times$10$^{-5}$ & 3.436$\times$10$^{-7}$ & 0.733 & 0.001 \\
$\rightarrow$ Endogenous minihalos & 1.429$\times$10$^{-5}$ & 4.323$\times$10$^{-7}$ & 0.717 & 0.001 \\
$\rightarrow$ Exogenous minihalos & 3.296$\times$10$^{-7}$ & 1.140$\times$10$^{-7}$ & 0.790 & 0.014 \\
LW w/ high star formation efficiency ($\epsilon_h$ = 0.2)* & 1.085$\times$10$^{-6}$ & 1.808$\times$10$^{-7}$ & 0.671 & 0.007 \\
$\rightarrow$ Endogenous minihalos & 2.469$\times$10$^{-6}$ & 4.734$\times$10$^{-7}$ & 0.620 & 0.008 \\
$\rightarrow$ Exogenous minihalos & 1.695$\times$10$^{-5}$ & 4.981$\times$10$^{-6}$ & 0.521 & 0.012 \\
LW w/ fiducial star formation efficiency ($\epsilon_0$ = 0.04)* & 1.191$\times$10$^{-6}$ & 1.262$\times$10$^{-7}$ & 0.688 & 0.004 \\
$\rightarrow$ Endogenous minihalos & 2.733$\times$10$^{-6}$ & 3.148$\times$10$^{-7}$ & 0.639 & 0.005 \\
$\rightarrow$ Exogenous minihalos & 4.050$\times$10$^{-6}$ & 1.020$\times$10$^{-6}$ & 0.595 & 0.010 \\
LW w/ low star formation efficiency ($\epsilon_l$ = 0.008)* & 1.497$\times$10$^{-6}$ & 9.628$\times$10$^{-8}$ & 0.708 & 0.003 \\
$\rightarrow$ Endogenous minihalos & 2.431$\times$10$^{-6}$ & 1.706$\times$10$^{-7}$ & 0.678 & 0.003 \\
$\rightarrow$ Exogenous minihalos & 3.380$\times$10$^{-6}$ & 6.301$\times$10$^{-7}$ & 0.615 & 0.008 \\
ACHs & 2.678$\times$10$^{-6}$ & 1.425$\times$10$^{-7}$ & 0.693 & 0.002 \\
ACHs \Vmax($z$ = 10) $\geq$ 30 km/s & 1.001$\times$10$^{-4}$ & 1.733$\times$10$^{-4}$ & 0.452 & 0.070 \\
\hline
\end{tabular}\\

* based on \citet{Crosby:2013bp}.
\end{center}
\label{tbl:param}
\end{table*}%

\subsubsection{Classical dwarfs and the Magellanic Clouds}
\label{sec:obs}

In Table~\ref{tbl:dwarfs}, we list the number of progenitor systems that we derived for a sample of nine classical dwarf galaxies. We used the abundance matching prescription of \citet{GarrisonKimmel:2014bca} to find the corresponding subhalo peak mass. We estimate that $\sim$154 atomic cooling halos were swallowed by the LMC prior to its infall into the Milky Way. Draco by comparison may have accreted only 10 atomic cooling halos by infall. Each column contains the number of estimated accreted minihalos which represent the total number of exogenous or endogenous minihalos which fell into the classical dwarf satellites prior to their own infall. The results are purely based on abundance matching \citep{GarrisonKimmel:2014bca} to infer the subhalo peak mass. Depending on the choice of star formation efficiency, we find that only a handful of star forming minihalos fell into the Draco system (1-3 exogenous systems and 3-5 endogenous systems) prior to infall. Direct treatment of the LW radiation at these early times will provide more solid estimates for the number of progenitor systems in each case.

\subsubsection{Ultra-faint dwarf galaxies}

The population of ancient, very low luminosity ``ultra-faint'' dwarf (UFD) galaxies in the Milky Way has been studied extensively for their star formation, chemical composition, and association to Galactic building blocks (see \citet{Frebel:2010kw} for a review). Recently, the Dark Energy Survey unveiled nine new such UFDs  (\citealt{desCollaboration:2015uo}, \citealt{Koposov:2015wb}). Interestingly, these satellites are close to the Large Magellanic Cloud (LMC). What remains to be answered, though, is whether most of the stellar material in such UFDs actually formed in-situ, or whether the dwarfs contain a substantial population of stars accreted from other, possibly chemically distinct, star forming systems.

This idea can in principle be tested with detailed chemical abundances of metal-poor stars that are found in all UFD galaxies. 
For example in the UFD Reticulum~II, seven of nine stars observed are strongly enhanced in heavy r-process elements which already led to the suggestion that this UFD experienced a massive r-process event by either a neutron-star merger or a jet driven supernova \citep{Ji:2015vc}. But the other two stars, which also happen to be the two most metal-poor stars in Reticulum~II, display extremely low abundances of those same heavy neutron-capture elements,  ([Ba/Fe] $<$ 0, \citealt{Ji:2015vc}). Furthermore, these nuclei were unlikely produced in an r-process but in some other event or site. 

These two groups of nucleosynthetic signatures suggest the following about the nature and evolution of Reticulum~II: either a) the stars with low heavy neutron-capture abundances formed within Reticulum~II but prior to the $r$-process enrichment event, or b) they formed in a pocket of low-metallicity gas that was not affected by the $r$-process enrichment. Importantly, the latter scenario could have occurred in a different, smaller system that was later accreted into Reticulum~II.

In general, our results (see Figure~\ref{fig:progenitors}) indicate that it is unlikely that many UFD candidates could have accreted more than a few (endogenous or exogenous) minihalos. The vast majority of potential progenitors were simply unable to form stars due to the H$_2$ dissociating by the onset of the LW background. Even under the most optimistic of circumstances where we assume a \cite{Tegmark:1997ce} minimum mass threshold for formation and remove our model for the LW background, the most massive of the future ultra-faint dwarf galaxies (\Mpeak $\sim$ 10$^{8.5}$ \Msol) accreted $<$10 minihalos.

With the inclusion of our fiducial LW model, this number is reduced to only one minihalo. Thus, the ``small system accretion'' scenario for Reticulum~II is unlikely and very few stars, if any, originate from distinct minihalos. However, larger dSphs like Draco and Ursa Minor are very likely to contain metal-poor stars from multiple progenitor minihalos. Moreover, detailed theoretical modelling of UFDs would greatly assist in this question also by constraining metal mixing and star formation processes to determine the exact origin of potentially different stellar abundance patterns within single UFD systems. Hydrodynamic simulations of UFDs may greatly assist interpretations of chemical abundances in UFDs by further constraining the impact of metal mixing and hierarchical galaxy formation on the exact origin of potentially different stellar abundance patterns within single UFD systems.

\begin{table*}
\caption{{Estimates of the number of progenitors for nine classical dwarf galaxies and Magellanic systems.}}
\begin{center}
\begin{tabular}{l|ccc|cccc|cccc|cccc|cccc}
\hline
\multirow{0}{*}{} &
      \multicolumn{1}{c}{} &
      \multicolumn{2}{c}{} &
      \multicolumn{15}{c}{Minihalos} \\

\multirow{0}{*}{Dwarf Galaxy} &
      \multicolumn{2}{c}{ACHs} &
      \multirow{0}{*}{} &
      \multicolumn{3}{c}{Tegmark} &
      \multirow{0}{*}{} &
      \multicolumn{3}{c}{$\epsilon_h$ = 0.2} &
      \multirow{0}{*}{} &
      \multicolumn{3}{c}{$\epsilon_0$ = 0.04} &
      \multirow{0}{*}{} &
      \multicolumn{3}{c}{$\epsilon_l$ = 0.008} \\
     
\multirow{0}{*}{} &
      \multicolumn{1}{|c}{$-$reion} &
      \multicolumn{1}{c|}{+reion} &
      \multirow{0}{*}{} &
      \multicolumn{1}{c}{En.} &
      \multicolumn{1}{c}{Ex.} &
      \multicolumn{1}{c|}{Total} &
      \multirow{0}{*}{} &
      \multicolumn{1}{c}{En.} &
      \multicolumn{1}{c}{Ex.} &
      \multicolumn{1}{c|}{Total} &
      \multirow{0}{*}{} &
      \multicolumn{1}{c}{En.} &
      \multicolumn{1}{c}{Ex.} &
      \multicolumn{1}{c|}{Total} &
      \multirow{0}{*}{} &
      \multicolumn{1}{c}{En.} &
      \multicolumn{1}{c}{Ex.} &
      \multicolumn{1}{c|}{Total}\\
\hline
Draco       & 12 & 0 && 120 & 7   & 127 && 1 & 0 & 1 && 3 & 1 & 4 && 8 & 1 & 9  \\
Ursa\ Minor & 12 & 0 && 120 & 7   & 127 && 1 & 0 & 1 && 3 & 1 & 4 && 8 & 1 & 9  \\
Carina      & 14 & 0 && 135 & 8   & 143 && 2 & 0 & 2 && 3 & 1 & 4 && 9 & 1 & 10  \\
Sextans\ I  & 14 & 0 && 143 & 9   & 152 && 2 & 1 & 3 && 4 & 1 & 5 && 9 & 1 & 10  \\
Leo\ II     & 18 & 0 && 178 & 13  & 191 && 2 & 1 & 3 && 5 & 1 & 6 && 11 & 2 & 13  \\
Sculptor    & 29 & 0 && 285 & 27  & 312 && 4 & 2 & 6 && 8 & 3 & 11&& 19 & 4 & 23  \\
Leo\ I      & 41 & 0 && 410 & 48  & 458 && 6 & 3 & 9 && 11& 4 & 15&& 27 & 6 & 33  \\
Fornax      & 71 & 2 && 706 & 111 & 817 && 12& 6 & 18&& 20& 9 & 29&& 47 & 12 & 59  \\
Sagitarrius & 72 & 2 && 723 & 115 & 838 && 12& 6 & 18&& 20& 9 & 29&& 48 & 13 & 61  \\
\hline
SMC         & 179& 20&& 1810& 473 & 2283&& 32& 19& 51&& 52& 29& 81&& 123 & 40 & 163  \\
LMC         & 246& 42&& 2503& 780 & 3283&& 46& 28& 74&& 72& 43&115&& 171 & 59 & 230  \\
\hline
\end{tabular}
\\
Note: ``En.'' represent endogenous systems and ``Ex.'' represent exogenous systems. ``$-$reion'' means no reionization included, ``+reion'' refers to the total number of atomic cooling halos which have \Vmax $\geq$ 30 km/s at $z$ = 10.
\end{center}
\label{tbl:dwarfs}
\end{table*}%

\subsection{Remnants of the first stellar systems in the Galaxy today}

With our 30 high-resolution simulations, we can quantify the halo-to-halo scatter in the remnant population. A full treatment requires more detailed modelling of the stellar mass associated with each remnant, but as a first step we tag the 10 of the most bound particles at $z$ = 10 for the minihalos identified with Lyman-Werner feedback at our fiducial star formation efficiency, $\epsilon_0$ = 0.04 (see Figure~\ref{fig:viz}) and determine their number density as a function of galactocentric distance.

\begin{figure}
\begin{center}
\includegraphics[width=0.48\textwidth]{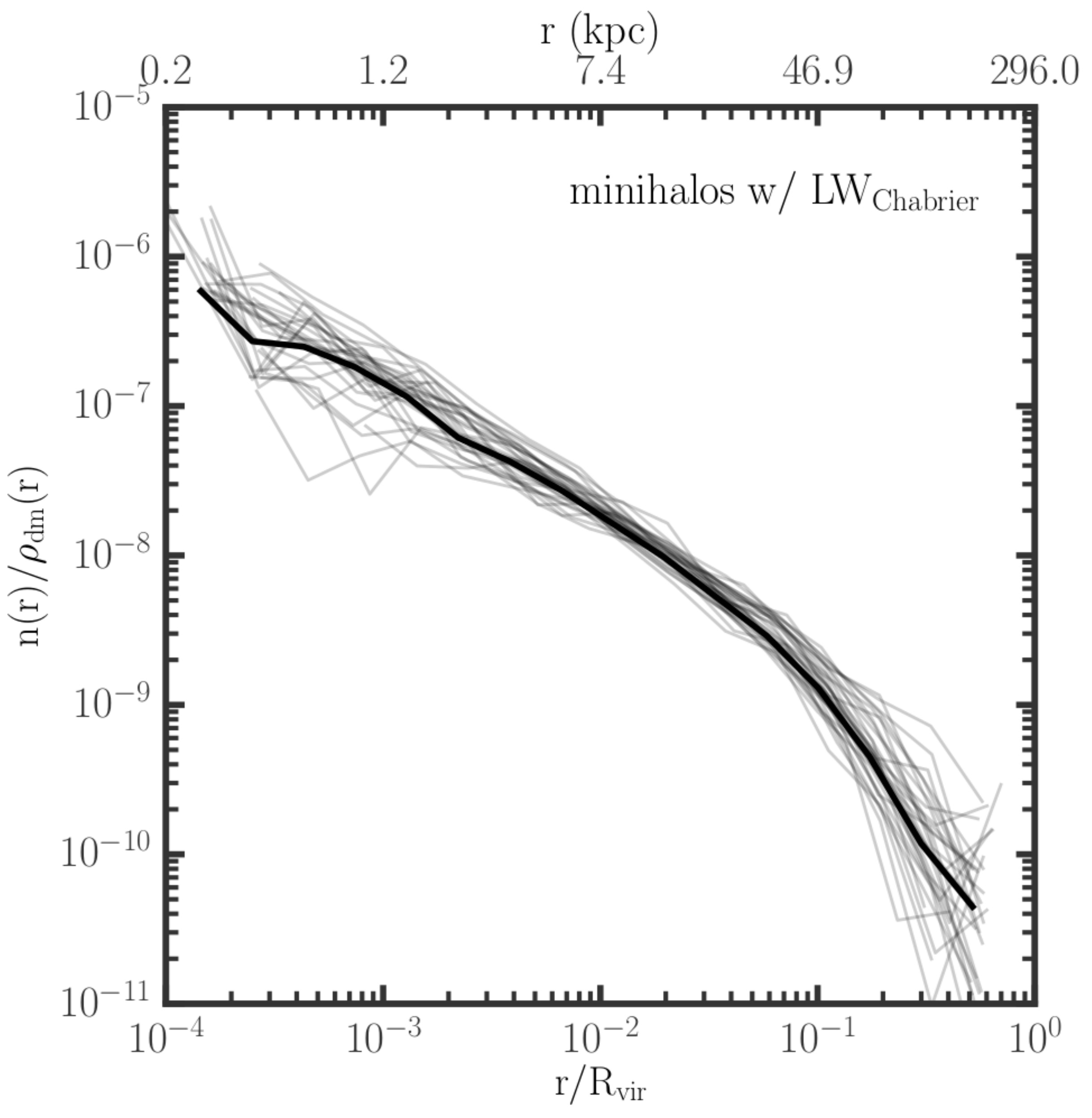}
\caption{{Ratio between radial number density profiles of Pop\, III remnants, $n(r)$, and dark matter mass densities of host halos ($\rho_\mathrm{dm}(r)$ for 30 \caterpillar halos (each individually marked in grey and median marked by thick black line). The top axis represents the bottom axis multiplied by the mean virial radius for all 30 runs (296\ kpc). The scatter in the number density at fixed galactocentric distance is quite large at small radii (e.g. an order of magnitude in the bulge) and large radii (i.e.\ r $>$ 50\ kpc) but similar within the halo (i.e.\ $r$ $<$ 30\ kpc). \label{fig:nr}}}
\end{center}
\end{figure}

In Fig.~\ref{fig:nr} we plot these number densities divided by the dark matter density of the host out to the virial radius for each host (black line is the median). This ratio highlights any bias in the remnant distribution relative to the overall density of particles in the dark matter halo. The scatter in the number density at fixed galactocentric distance is an order of magnitude at small radii (e.g.\ within the bulge) and large radii (i.e.\ r $>$ 50\ kpc) but similar within the halo (i.e.\ r $<$ 30\ kpc). Our scatter agrees qualitatively with the result found by \cite{Ishiyama:2016vt} who used four halos. Additionally, we find different overall means owing to the alternative Lyman-Werner treatment and slightly different tagging method (i.e\ at formation versus at $z$ = 10). \cite{Tumlinson:2009ia} has argued that metal-poor stars in the bulge are most likely to be true relics of Pop II. stars. However, \cite{Salvadori2015} and more recently, \cite{Starkenburg:2016arxiv} find that the oldest stars populate the innermost region of the Galaxy while the relative contribution of very metal poor stars increases with radius from the Galactic center. Without more detailed modeling, we can not compare directly with these works except to state that our oldest remnants populate all parts of the Galaxy with scatter most pronounced in the bulge and at large radii.

\section{Conclusions}
\label{sec:conclusions}

We have presented a systematic study of the general properties of minihalos and atomic cooling halo progenitors of Milky Way sized systems using 30 cosmological simulations. In our model for first star formation, we include the impact of Lyman-Werner radiation on the earliest stellar systems and determine how the clustering properties of such star-forming systems enriched subsequent generations of stars and galaxies in the Milky Way. Our model and results can be summarized as follows, first with respect to minihalos then atomic cooling halos.

\subsection{Minihalo Progenitors of Milky Way sized systems}
Using a physically motivated minimum mass threshold, we identify all molecular line cooling halos via the cooling threshold of \citet{Tegmark:1997ce}. We additionally use the semi-analytic prescriptions of \citet{Crosby:2013bp} for the LW background using three different star formation efficiencies ($\epsilon$ = 0.008, 0.04, 0.2). We find the following:
\begin{itemize}
\item Without LW feedback, we find 22,856 $\pm$ 4915 progenitor dark matter halos of a Milky way sized host to satisfy the minimum mass threshold required for their molecular hydrogen gas to cool, collapse and form stars.
\item With LW feedback, the number of potential star forming minihalo progenitors is significantly reduced (by $\sim$90$\%$) because the radiation raises the minimum mass required to form stars. We find  358$\pm$82/653$\pm$141/1458$\pm$314 (for star formation efficiencies: $\epsilon_l$ = 0.008, $\epsilon_0$ = 0.04, $\epsilon_h$ = 0.2) minihalos satisfy our requirements to form stars and eventually merge into the host halo.
\item By $z$ = 0, 55$\%$ of all progenitor systems are accreted by the central host and the remainder reside within subhalos of the central host.
\item Using a simple chemical enrichment model, we determined what fraction of systems have their chemical composition established by in-situ star formation or by being enriched by neighboring systems. Overwhelmingly, most of the minihalos evolve in isolation without the influence (chemically) of an external halo (i.e.\ 80-90$\%$ of all systems at $z$ = 7 are endogenous). For the strong feedback model, we find $\sim$18$\%$ of systems are exogenous at $z$ = 7 compared to $<$7$\%$ of systems for the weak and fiducial feedback models. When halos are externally enriched, it is usually by 10$^8$ \Msol systems or more massive ones.
\item  Of the systems which are endogenous, $\sim$50$\%$ merge with a system larger than themselves within 1.5 Gyr after formation (Fig. \ref{fig:mergetimescales}). This leads to enhanced chemical enrichment, making them only temporarily endogenous systems. Several generations of stars could have formed (and died) between the time of first star formation and the eventual accretion of the system into the main host.
\item Star forming minihalos are on average median separated to other star forming 10$^7$ \Msol halos by 300\,pc at $z$ = 20 and 3\,kpc at $z$ = 7. While we found most systems are internally enriched, a more realistic chemical enrichment model including proper treatment of chemical mixing and non-instantaneous winds may result in an increase in the externally enriched fraction.
\item The number of minihalo progenitor systems which have been accreted by a given subhalo halo is best fit via a power law. The number of exogenous progenitors is best fit via the power law, $N_{\mathrm{prog}}\ =\ 2.97\ \times\ 10^{-4} M_{\mathrm{peak}}^{0.4}$. Similarly, the number of endogenous progenitors for a given subhalo is best fit by $N_{\mathrm{prog}}\ =\ 4.82\ \times\ 10^{-7} M_{\mathrm{peak}}^{0.71}$. 
\item We estimate that there is an order of magnitude scatter in the number density of Pop\, III remnants at small (i.e.\ r $<$ 5\ kpc) and large galactocentric radii (i.e.\ r $>$ 50\ kpc) across Milky Way-mass halos. The scatter is most minimal at intermediate distances (10 $<$ r $<$ 50\ kpc) within the halo.
\item We estimate that low luminosity UFD galaxies, such as Reticulum~II, have at most one or two star forming minihalo progenitors. Consequently, it highly unlikely that Reticulum~II received its $r$-process enriched material via an external system bringing in chemically enriched stellar material.
\item Similarly, we estimate that approximately $\sim$74-230 ($\sim$51-163) minihalos were accreted by the proto-LMC (SMC), creating a potentially large number of ultra-faint satellite systems which could be tidally removed from the LMC during first passage and distributed throughout the Milky Way.
\end{itemize}

\subsection{Atomic Cooling Halo Progenitors of Milky Way sized systems}
We identified all potentially atomic cooling halos in each simulation and used a simple model of reionization to determine which halos were suppressed, partially suppressed and active in the post-reionization era ($z$ $<$ 10). Our results can be summarized as follows:

\begin{itemize}
\item There are 1793 $\pm$ 396 atomic cooling halo progenitors per 10$^{12}$ \Msol host (across 30 Milky Way sized systems).
\item We find 781 $\pm$ 215 (44 $\pm$ 12$\%$) of these systems do not survive the reionization era and will stop accreting gas and forming new stars (937 additional systems form after $z$ = 10 with \Vmax $<$ 30 km/s and are suppressed in our model). 
On average, we also also find that 64 (4 $\pm$ 2$\%$) systems are partially suppressed and 11 (1 $\pm$ 1$\%$) systems are not suppressed at all and will continue to accrete gas and form stars unimpeded by reionization. These will accrete into either larger progenitors and become dwarf galaxies or be disrupted during the accretion onto the primary host.
\item By z$\sim$0, 54$\%$ of the unsuppressed atomic cooling halo progenitor systems are accreted by the central host and the remainder end up within the subhalos.
\item The number of atomic cooling halo progenitor systems of a given subhalo of the host is best fit via the power law, $N_{\mathrm{prog}}\ =\ 2.69\ \times\ 10^{-7} M_{\mathrm{peak}}^{0.69}$.
\item Approximately 246 atomic cooling halos were accreted by the LMC prior to infall and $\sim$12 atomic cooling halos were accreted by Draco. Using a simple model for reionization, we find only 42 (0) of these LMC (Draco) progenitor systems have \Vmax($z$ = 10) $\geq$ 30 km/s and will survive the reionization era.
\end{itemize}

We finally comment that \cite{Gao:2010cva} used the $Aquarius$ simulation suite to identify Pop\,III star forming progenitors. They employed a similar method as ours, though at a lower virial temperature threshold (1100\,K). They found $\sim$2$\times$10$^4$ Pop\,III star forming progenitors which agrees well with our estimates of $\sim$23,000. Similarly, they found a mean separation distance of $\sim$1\ h$^{-1}$\,kpc ($z$ = 10) which also agrees well with our estimates ($\sim$3\ h$^{-1}$\,kpc). They also determine the number of first galaxies (i.e.\ 10$^4$\,K) to be $\sim$200-300 by $z$ = 10. We speculate that this estimate is lower than ours because of the lower temporal resolution used in the $Aquarius$ simulation suite ($\sim$100\,Myrs/snapshot outputs compared to $\sim$ 5\,Myr/snapshot outputs in \caterpillar). We use a different model for the LW background than the work of \cite{Ishiyama:2016vt} and so it make it difficult to compare numbers directly. Additionally, \cite{Gao:2010cva} also do not provide population statistics which furthermore complicates a detailed comparison of results.

\subsection{Caveats $\&$ Future Work}
\label{sec:caveats}

Our modelling technique is not without drawbacks. Most importantly, we do not resolve the direct collapse of gas, subsequent fragmentation and enrichment \textit{directly} and rely on the assumption that a given halo's temperature is in virial equilibrium with the gas temperature. We additionally assume that the enrichment process proceeds via instantaneous, spherically enriched gas bubbles at a scale set purely by the progenitor host halo mass. It is known from detailed hydrodynamic simulations of single halo systems that star formation proceeds in a much more stochastic manner and that the enrichment process is very unstructured and depends heavily on local environmental conditions. Despite these limitations, we are providing a robust machinery for connecting present day halos with their high-z progenitors, and offer a first glimpse to statistically probe the locations of the first star forming progenitors of Milky Way-mass halos by sampling the largest number of Milky Way halos ever simulated at such high resolution.

The results of this work will invite more direct semi-analytic modelling of the relevant star formation and feedback processes in the future. Moving forward, we aim to more self-consistently model the formation sites of the first stellar systems and subsequent first galaxies including an enhanced treatment of the relevant radiative processes crucial to regulating each progenitor's assembly history. This modelling will then allow a more detailed understanding of the origin of the chemical make-up of not only the old stellar halo, but also its satellite systems. Only self-consistently modelling of the chemical and dynamical evolution of all of the progenitors of a Milky Way sized host will enable theoretical progress capable of connecting the low-redshift universe to the earliest phases of galaxy formation. Coupling the rich chemical and kinematic data being released by various observational Galactic sky surveys (e.g., GAIA-ESO, Gaia, SkyMapper, GALAH) with advanced modelling of this kind will contribute significantly to the nascent areas of both stellar and dwarf galaxy archaeology.

\section*{Acknowledgements}
BG would like to thank Paul Hsi for assistance with the compute cluster at MKI. He would also like to thank Bhaskar Agarwal, Andy Casey and Joss Bland-Hawthorn for helpful discussions. The authors thank Oliver Hahn for making the initial conditions code, {\sc{music}}, publicly available. The authors also thank Volker Springel for making {\sc{gadget-2}} publicly available and for providing a version of {\sc{gadget-3}}/{\sc{gadget-4}} for our use. The authors thank Peter Behroozi for making {\sc{rockstar}} and {\sc{consistent-trees}} publicly available and additionally thank him for technical support in modifying {\sc{rockstar}}.

Support for this work was provided by XSEDE through the grants (TG-AST120022, TG-AST110038). BG and AF acknowledges support of the compute cluster of the Astrophysics Division which was built with support from the Kavli Investment Fund administered by the MIT Kavli Institute for Astrophysics and Space Research. GD acknowledges support by NSF Grant 1122374. BWO and FG were supported through the NSF Office of Cyberinfrastructure by grant PHY-0941373 and by the Michigan State University Institute for Cyber-Enabled Research (ICER).  BWO was supported in part by NSF grant PHY 08-22648 (Physics Frontiers Center/Joint Institute for Nuclear Astrophysics) and NSF Grant  PHY-1430152 (JINA Center for the Evolution of the Elements), and by NASA through grants NNX12AC98G, NNX15AP39G, and Hubble Theory Grants HST-AR-13261.01-A and HST-AR-14315.001-A.  AF acknowledges support from the Silverman (1968) Family Career Development professorship.

\bibliographystyle{mnras}
\bibliography{bibliography}

\section{Appendix}

\begin{figure}
\begin{center}
\includegraphics[width=0.98\columnwidth]{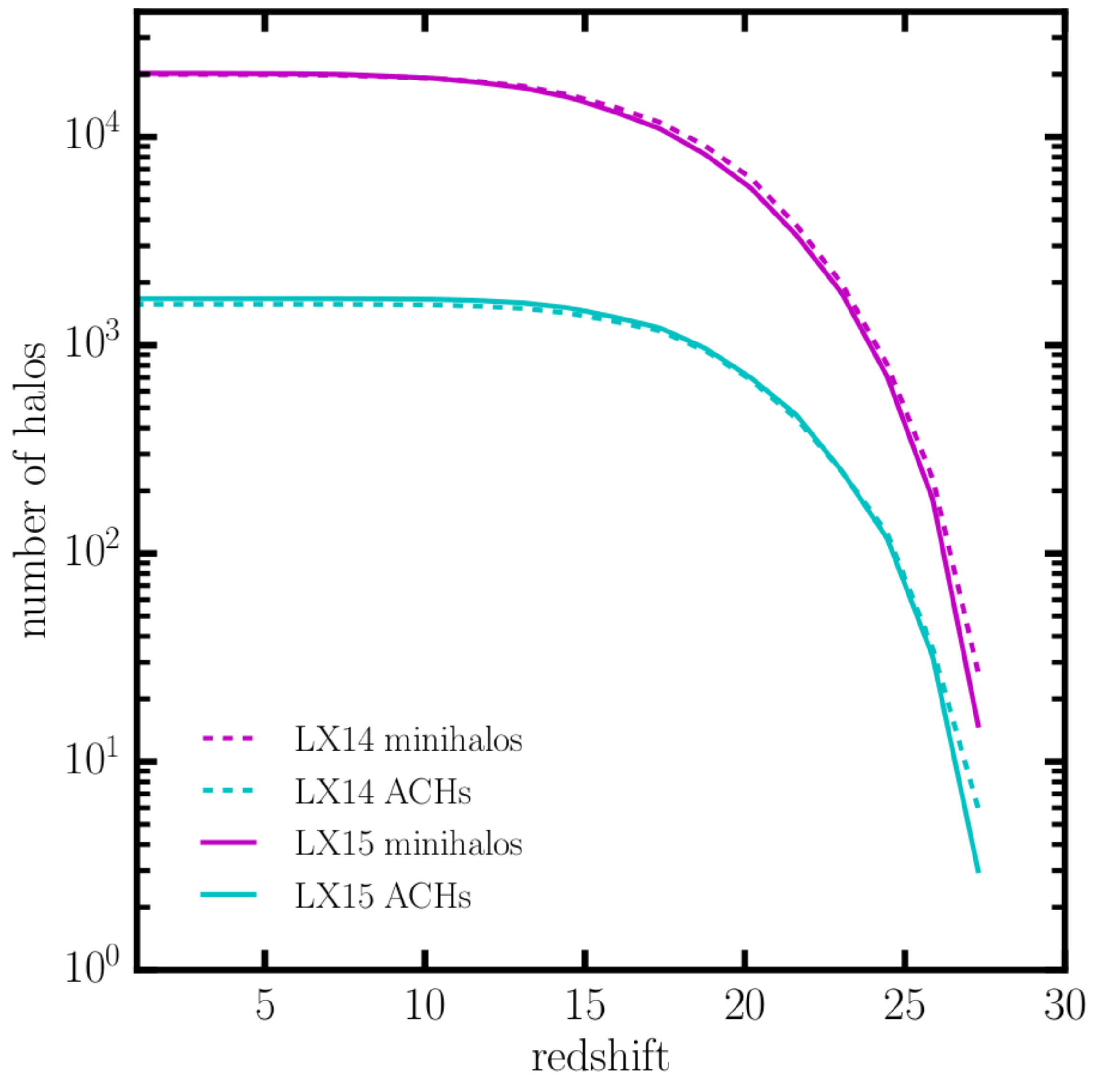}
\caption{{The total number of halos identified as molecular line cooling halos (minihalos) and atomic cooling halos in both our fiducial run (LX14, $m_p$ =  2.99 $\times$ 10$^4$   \Msol) and a higher resolution run (LX15, $m_p$ = 3.73 $\times$ 10$^3$ \Msol).
\label{fig:convergence}%
}}
\end{center}
\end{figure}

We also carried out the same analysis on a higher resolution halo (LX15, $m_p$ = 3.73 $\times$ 10$^3$ \Msol) which has a particle mass eight times higher than our fiducial run (LX14, $m_p$ = 2.99 $\times$ 10$^4$ \Msol) to check that we identify the same total number of systems. The total number of systems identified as atomic cooling halos and molecular line cooling halos (minihalos) in the two resolution runs of the Cat-9 halo are shown in Figure \ref{fig:convergence}. We find good agreement between the runs.

\end{document}